\documentclass[twocolumn,showpacs,preprintnumbers,amsmath,amssymb]{revtex4}
\usepackage{graphicx}% Include figure files
\usepackage{dcolumn}% Align table columns on decimal point
\usepackage{bm}% bold math

\begin{document}

\preprint{APS/123-QED}

\title{Evaluation of mutual information estimators on nonlinear dynamic systems}

\author{Angeliki Papana}
\email{agpapana@gen.auth.gr}
\author{Dimitris Kugiumtzis}
\email{dkugiu@gen.auth.gr}
\affiliation{Department of
Mathematical, Physical and Computational Sciences, Faculty of
Engineering, Aristotle University of Thessaloniki, Thessaloniki
54124, Greece }

\date{\today}% It is always \today, today,

\begin{abstract}
Mutual information is a nonlinear measure used in time series
analysis in order to measure the linear and non-linear
correlations at any lag $\tau$. The aim of this study is to
evaluate some of the most commonly used mutual information
estimators, i.e. estimators based on histograms (with fixed or
adaptive bin size), $k$-nearest neighbors and kernels. We assess
the accuracy of the estimators by Monte-Carlo simulations on time
series from nonlinear dynamical systems of varying complexity. As
the true mutual information is generally unknown, we investigate
the existence and rate of consistency of the estimators
(convergence to a stable value with the increase of time series
length), and the degree of deviation among the estimators. The
results show that the $k$-nearest neighbor estimator is the most
stable and less affected by the method-specific parameter.
\end{abstract}

\pacs{Valid PACS appear here}% PACS, the Physics and Astronomy

\maketitle

%=========================================
\section{Introduction}
%=========================================
Mutual information is a popular nonlinear measure of time
series analysis, best known as a criterion to select the
appropriate delay for state space reconstruction
\cite{Kantz97}. It is also used to discriminate different
regimes of nonlinear systems \cite{Chillemi03,Wicks07} and to
detect phase synchronization \cite{Schmid04,Kreuz07}. Besides
nonlinear dynamics, it is used in various statistical settings,
mainly as a distance or correlation measure in data mining,
e.g. in independent component analysis and feature-based
clustering \cite{Tourassi01,Priness07}.

It is well-known that any estimate of mutual information, either
between two variables or as a function of delay for time series,
is (almost always) positively biased
\cite{Treves95,Moddemeijer89,Paninski03}. For numerical-valued
variables and time series, the mutual information increases with
finer partition depending on the underlying distribution or
process and the sample size. Beyond the classical histogram-based
partitioning, other schemes have been used to estimate the
densities inherent in the measure of mutual information, e.g.
using kernels and $k$-nearest neighbors \cite{Moon95,Kraskov04}.

Although there are some works comparing mutual information
estimators in
\cite{Moon95,Darbellay99,Steuer02,Daub04,Kraskov04,Cellucci05,Nicolaou05,Trappenberg06,Khan07},
to the best of our knowledge, there has not been a comparison of
all commonly used estimators, including the selection of their
parameters, on time series from dynamical deterministic systems.

Moon et al. \cite{Moon95} developed a kernel mutual information
estimator, as an extension of Silverman's work \cite{Silverman86}.
This estimator is compared to the locally adaptive histogram-based
estimator of Fraser and Swinney \cite{Fraser86} on four linear and
nonlinear systems using as a performance criterion the lag of the
first minimum of mutual information. Darbellay and Vajda
\cite{Darbellay99} suggested an adaptive histogram-based estimator
and compared it with mutual information estimators derived from
maximum likelihood estimators for some bivariate distributions
with analytically known mutual information. Steuer et al.
\cite{Steuer02} presented three histogram-based estimators,
investigated their bias and suggested using the kernel estimator.
Daub et al. \cite{Daub04} estimated mutual information using
B-spline functions and compared it to an entropy estimator
suggested by Paninski \cite{Paninski03} and a kernel density
estimator on data sets drawn from a known distribution. They
claimed that their method is computationally faster than the
kernel estimator and improves the simple binning method. Kraskov
et al. \cite{Kraskov04} developed an estimator of mutual
information based on $k$-nearest neighbors and compared it to the
adaptive histogram-based estimator of Darbellay and Vajda but only
on Gaussian and some non-Gaussian distributions with analytically
known mutual information. In \cite{Cellucci05}, equidistant and
equiprobable histogram-based estimators (using three selection
criteria for the number of bins $b$) are compared to the algorithm
of Fraser and Swinney on nonlinear systems as to their robustness
in detecting a fixed delay for the first minimum of the mutual
information (similarly to \cite{Moon95}). They also use the bias
as a performance criterion in the case of Gaussian processes
(where the true mutual information is known) and find that the
equiprobable histogram-based estimator is more accurate and the
Fraser and Swinney estimator is computationally ineffective.

In a different setting, in \cite{Nicolaou05} the estimators of
equidistant histograms, kernels, B-splines, and $k$-nearest
neighbors, are tested on electroencephalographic data from rats in
order to find dependencies between left and right channels. Using
the surrogate data test for the significance of dependence and
bootstrap confidence intervals for the estimators, they concluded
that the B-spline estimator is largely affected by its parameter
and the $k$-nearest neighbor is the most consistent and less
dependent on its parameter. Trappenberg et al.
\cite{Trappenberg06} compared the equidistant histogram-based
method, the adaptive histogram-based method of Darbellay and Vajda
and the Gram-Charlier polynomial expansion \cite{Blinnikiov98} and
concluded that all three estimators gave reasonable estimates of
the theoretical mutual information, but the adaptive
histogram-based method converged faster with the sample size. A
more comprehensive evaluation of mutual information estimators
including the kernel, $k$-nearest neighbor, equiprobable
histogram-based estimators and an estimator using the Edgeworth
approximation to estimate densities, was recently presented in
\cite{Khan07}, focusing on the deviation of the mutual information
from a linear correlation measure on linear and nonlinear time
series (using the Henon map for the chaotic case). In the same
paper, a small scale simulation showed dependence of the
performance of the kernel and nearest neighbor estimators on their
parameter.

Given the varying results in the literature on the different
mutual information estimators, we study here the performance of
three commonly used estimators, i.e. estimators based on
histograms (with fixed or adaptive bin size), $k$-nearest
neighbors and kernels, on time series from nonlinear deterministic
systems. Moreover, we investigate the optimal parameter for the
determination of the two-dimensional partitioning for each method.
Monte-Carlo simulations of dynamical systems of varying complexity
and observational noise level are used in order to assess the
accuracy of the estimators. Commonly used parameter selection
methods are considered for all but the kernel estimators, where a
range of bandwidths are tested.

The structure of the paper is as follows. In
Section~\ref{sec:Estimators} the estimators considered in this
study are presented. In Section~\ref{sec:Simulations} the results
of the simulations are presented and in
Section~\ref{sec:Discussion} the results are discussed and
conclusions are drawn.

%=========================================
\section{Estimators of mutual information}
\label{sec:Estimators}
%=========================================

In information theory, mutual information is defined as a measure of mutual
dependence of two variables $X$ and $Y$ and has the form
\begin{equation}
 \mathbf{I}(X,Y) = \int_X \int_Y f_{X,Y}(x,y) \log_a {\frac{f_{X,Y}(x,y)}
{f_X(x)f_Y(y)}}\mbox{d}x\mbox{d}y
 \label{eq:def1}
\end{equation}
where $f_{X,Y}(x,y)$ is the joint probability density function
(pdf) of $X$ and $Y$, and $f_X(x)$, $f_Y(y)$ are the marginal pdfs
of $X$ and $Y$, respectively. The units of information of $I(X,Y)$
depend on the base of the logarithm, e.g. bits for the base of
$2$ and nats for the natural logarithm in (\ref{eq:def1}).

Assuming a partition of the domain of $X$ and $Y$ the double
integral becomes a sum over the cells of the two-dimensional
partition:
\begin{equation}
 \mathbf{I}(X,Y) = \sum_{i,j}p_{X,Y}(i,j) \log_a {\frac{p_{X,Y}(i,j)}{p_X(i)p_Y(j)}}
 \label{eq:def2}
\end{equation}
where $p_X(i)$, $p_Y(j)$, and $p_{X,Y}(i,j)$ are the marginal and
joint probability distributions over the elements of the
partition. In the limit of fine partitioning the expression in
\eqref{eq:def2} converges to \eqref{eq:def1}. This may partly
justify the abuse of notation of mutual information for the
continuous and the discretized variables.

It is always $\mathbf{I}(X,Y) \ge 0$, with equality holding for
independent variables, and $\mathbf{I}(X,Y) \le H(X) \le
\log_a{n}$ (Jensen inequality), where $H(X)= -\sum_{i=1}^n
p(x_i) \log_a {{1}/{p(x_i)}}$ is the entropy of $X$. We do not
discuss the mutual information in terms of entropies as the
estimation of mutual information we study in this work boils
down to the estimation of the densities in \eqref{eq:def1} or
probabilities in \eqref{eq:def2}.

For a time series $\{X_t\}_{t=1}^n$, sampled at fixed times
$\tau_s$, the mutual information is defined as a function of
the delay $\tau$ assuming the two variables $X=X_t$ and
$Y=X_{t-\tau}$, i.e.
$\mathbf{I}(\tau)=\mathbf{I}(X_t,X_{t-\tau})$.

The true mutual information is generally not known as joint and
marginal probability density functions are unknown. A different
estimator $I(\tau)$ of $\mathbf{I}(\tau)$ is determined from
the way the theoretical densities in \eqref{eq:def1} or
probabilities in \eqref{eq:def2} are estimated. We discuss
below three estimators considered in this work and their
dependency on a parameter inherent in the estimation of the
densities or probabilities.

\subsection{Histogram-based estimators}
%======================================
The naive histogram-based estimator regards a partition of the
range of values of each variable into $b$ discrete bins of equal
length, termed as {\em equidistant partitioning} (ED). The density
at each bin and each two-dimensional cell, or rather the
probability functions in \eqref{eq:def2}, are estimated by the
corresponding relative frequency of occurrence of samples in the
bin or cell. Many different criteria have been developed for the
selection of the number of bins $b$ or equivalently the length of
each bin, e.g. see \cite{Bendat66,Duda73,Scott92,Knuth06}.
Alternatively, the partition can be done into equiprobable bins,
so that each bin has the same occupancy, termed as {\em
equiprobable partitioning} (EP). In any case the partitioning is
the same for both variables and the only free parameter is $b$. A
number of criteria for selecting $b$ for one-dimensional binning
can been used and in \cite{Cellucci05} the Cochran condition
(requiring at least 5 samples in a bin) extended for
two-dimensional cells is used to select $b$.

An extension of the equidistant and equiprobable partitioning is
the adaptive partitioning of the two-dimensional plane. Darbellay
and Vajda built an algorithm to estimate the mutual information
(AD) by calculating relative frequencies on appropriate partitions
formed in a way that conditional independence is achieved on the
cells \cite{Darbellay99}. The advantage of the estimator of
Darbellay and Vajda is that it is adaptive to the data and does
not involve a parameter for the binning. It does however involve a
parameter for the independence test that can affect the
performance of the estimator.

\subsection{$k$-nearest neighbor estimator}
%========================================
Kraskov et al. proposed an estimator of mutual information that
uses the distances of $k$-nearest neighbors to estimate the joint
and marginal densities \cite{Kraskov04} (KNN). For each reference
point from the bivariate sample, a distance length is determined
so that $k$ neighbors are within this distance length. Then the
number of points with distance less than half of this length give
the estimate of the joint density at this point and the respective
neighbors in one-dimension give the estimate of the marginal
density for each variable. The algorithm uses discs (or squares
depending on the metric) of a size adapted locally and then uses
the corresponding size in the marginal subspaces, so in some sense
the estimator is data adaptive. However, it involves as a free
parameter the number of neighbors $k$; large $k$ regards a small
$b$ of the histogram-based estimator. However, the estimator does
not use a fixed neighborhood size and therefore there is not a
clear association of $k$ and $b$.

\subsection{Kernel estimator}
%=============================
The kernel density estimator constructs a smooth estimate of the
unknown density by centering kernel functions at the data samples;
kernels are used to obtain the weighted distances
\cite{Silverman86,Moon95} (KE). The kernels essentially weight the
distances of each point in the  sample to the reference point
depending on the form of the kernel function and according to a
given bandwidth $h$, so that small $h$ produces more detail in the
density estimate. Thus $h$ plays the role in the kernel estimator
that $b$ plays in the histogram-based estimator (e.g. a
rectangular kernel assigns a histogram) and actually $h$ has an
inverse relation to $b$. It's advantage over histogram-based
estimators is that it is independent of the location of the bins.
Among the different kernel functions, Gaussian kernels are most
commonly used and we use them here as well. This estimator
involves actually two free parameters: the bandwidth $h_1$ for the
estimation of the marginal densities and the bandwidth $h_2$ for
the estimation of the joint density. Kernel estimators are
considered to be the most appropriate for density estimation of
one-dimensional data, but this does not necessary imply that the
kernel estimator of mutual information is also the most
appropriate.

%================================================
\section{Simulations and results}
\label{sec:Simulations}
%================================================

\subsection{Simulation setup}
%============================
The evaluation of the estimators is assessed by Monte-Carlo
simulations on the following chaotic systems: Henon and Ikeda
map, and Mackey Glass differential system with delay $\Delta =
17, 30, 100$ regarding increasing complexity with $\Delta$ (the
sampling time is $17\mbox{s}$). The factors considered are the
time series length $n$, given in a power of 2 from 8 to 13, and
the noise level, i.e. the standard deviation of additive
Gaussian noise is $20\%, 40\%$ and $80\%$ of the standard
deviation $s$ of the data. ${\it I}(\tau)$ is computed using
all methods on $1000$ realizations from the above systems up to
a lag $\tau$ for which ${\it I}(\tau)$ converges to a
non-negative constant value.

For each method, the corresponding free parameter covers a wide
range of values. For the ED and EP estimators we set the number of
bins to be $b = 2, 4, 8, 16, 32, 64$. The same values are set for
the parameter $k$ of the KNN estimator.

For the KE estimator we take $15$ different values of $h_1$ from
$0.01$ to $2$ with increment in logarithmic scale and $h_2=h_1$ or
$h_2=\sqrt{2}h_1$ (in order to account for the increase of
distance from one to two dimensions using the Euclidean norm).
Note that the values of $h$ regard the standardized data. We also
consider some well-known criteria for the selection of bandwidth
given in Table~\ref{tab:bandwidth}. The three first criteria
define bandwidth for both one and two-dimensional space. For the
last three criteria we set the two-dimensional bandwidth $h_2$ to
be either equal to $h_1$ or multiplied with $\sqrt{2}$.

\begin{table}[h!]
\caption{Criteria for the selection of bandwidths for one
($h_1$) and two ($h_2$) dimensions and the reference of the
source (last column). The parameters in the expressions are
$a=1.8-r(1)$ if $n<200$ and $a = 1.5$ if $n>=200$, where $r(1)$
is the autocorrelation at lag $1$, $R =  1/2{\sqrt{\pi}}$,
$\mbox{IQ}$ is the interquartile range of the data.}
\begin{tabular}{|c|l|l|l|} \hline
  & $h_1$ & $h_2$ & Ref \\ \hline
C1 & $(4/3n)^{(1/5)}$ &$(1/n)^{(1/6)}$ &\cite{Silverman86}\\\hline
C2 & $(4/3n)^{(1/5)}$ &$(4/5n)^{(1/6)}$&\cite{Silverman86}\\\hline
C3 & $1.06an^{(-1/5)}$ & $an^{(-1/6)}$ & \cite{Harrold01} \\\hline
C4 &   &   $h_1$   &   \\  \cline{1-1} \cline{3-3}
C5&\raisebox{2mm}[0pt]{$(8{\sqrt{\pi}}R/3n)^{(1/5)}\min(s,\mbox{IQ}/1.349)$}
&$\sqrt{2}h_1$&\raisebox{2mm}[0pt]{\cite{Silverman86,Wand95}}\\\hline
C6 &    & $h_1$ &   \\ \cline{1-1} \cline{3-3}

C7 & \raisebox{2mm}[0pt]{L-stage direct plug-in} & $\sqrt{2}h_1$ &
\raisebox{2mm}[0pt]{\cite{Wand95}}\\\hline

C8 &     & $h_1$  & \\ \cline{1-1} \cline{3-3}
C9 & \raisebox{2mm}[0pt]{Solve-the-equation plug-in} & $\sqrt{2}h_1$ & \raisebox{2mm}[0pt]{\cite{Sheather91}}\\
\hline
\end{tabular}
 \label{tab:bandwidth}
\end{table}

\subsection{Evaluation criteria}
%===============================

${\bf I}(\tau)$ is generally not known for non-linear chaotic
systems. In order to evaluate the mutual information
estimators, we examine their consistency and their dependence
in the corresponding parameters for all systems and time series
lengths. Regarding consistency, an estimator is consistent if
it converges to the true value with $n$. In the lack of the
true mutual information, we use as a reference value the ${\it
I}(\tau)$ computed on a realization of length $n=10^7$ for each
system (for some systems and estimators we considered $n=10^5$
or $n=10^6$ for time consuming reasons). Specifically, for time
series from the flows of the Mackey Glass systems we focus also
on the first minimum of the estimated mutual information and
examine the dependence of this on the free parameter of the
estimator and the time series length.

\subsection{Equidistant estimator}
\begin{figure}[h!]
\centerline{\hbox{
\includegraphics[height=3.4cm]{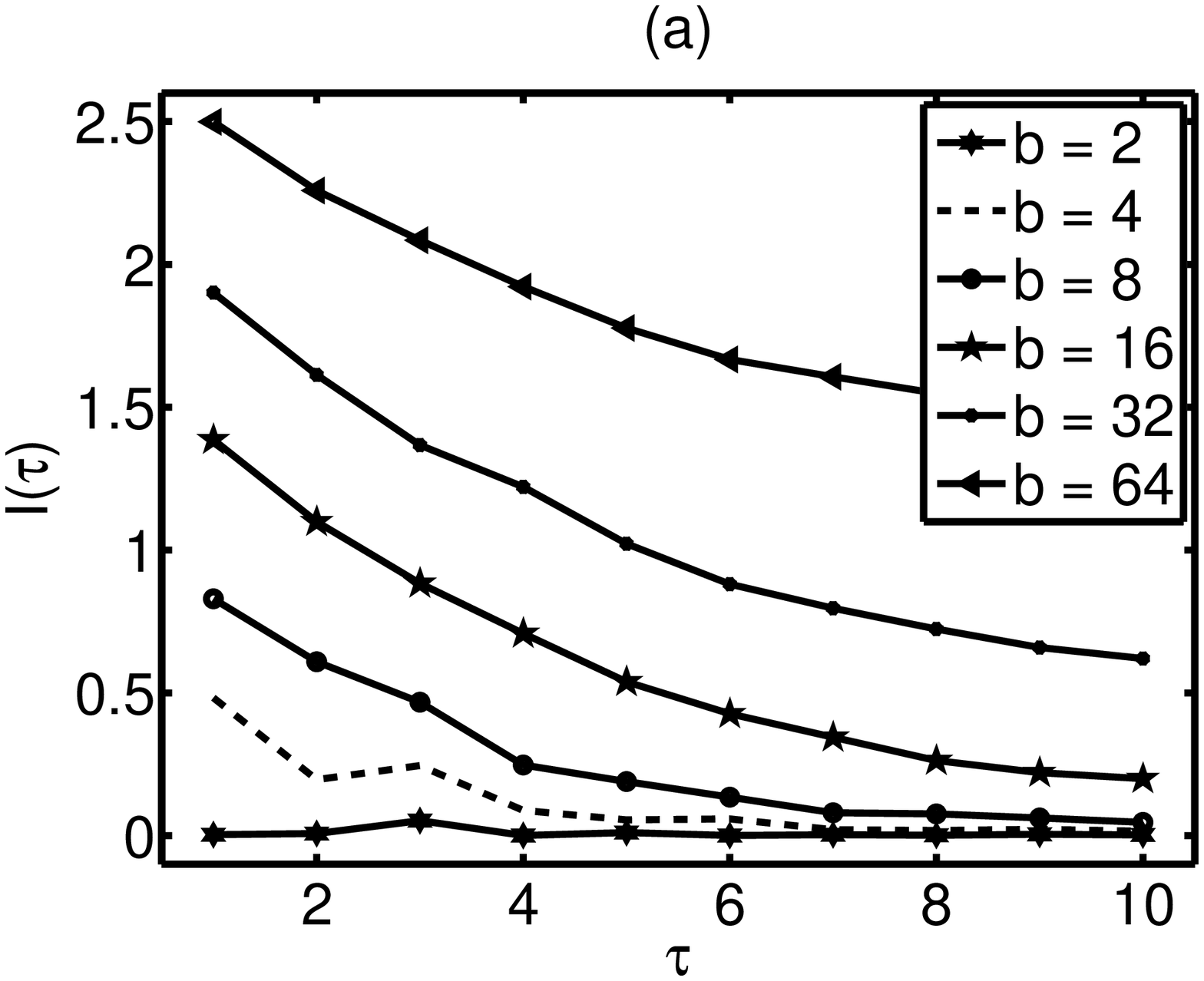}
\includegraphics[height=3.4cm]{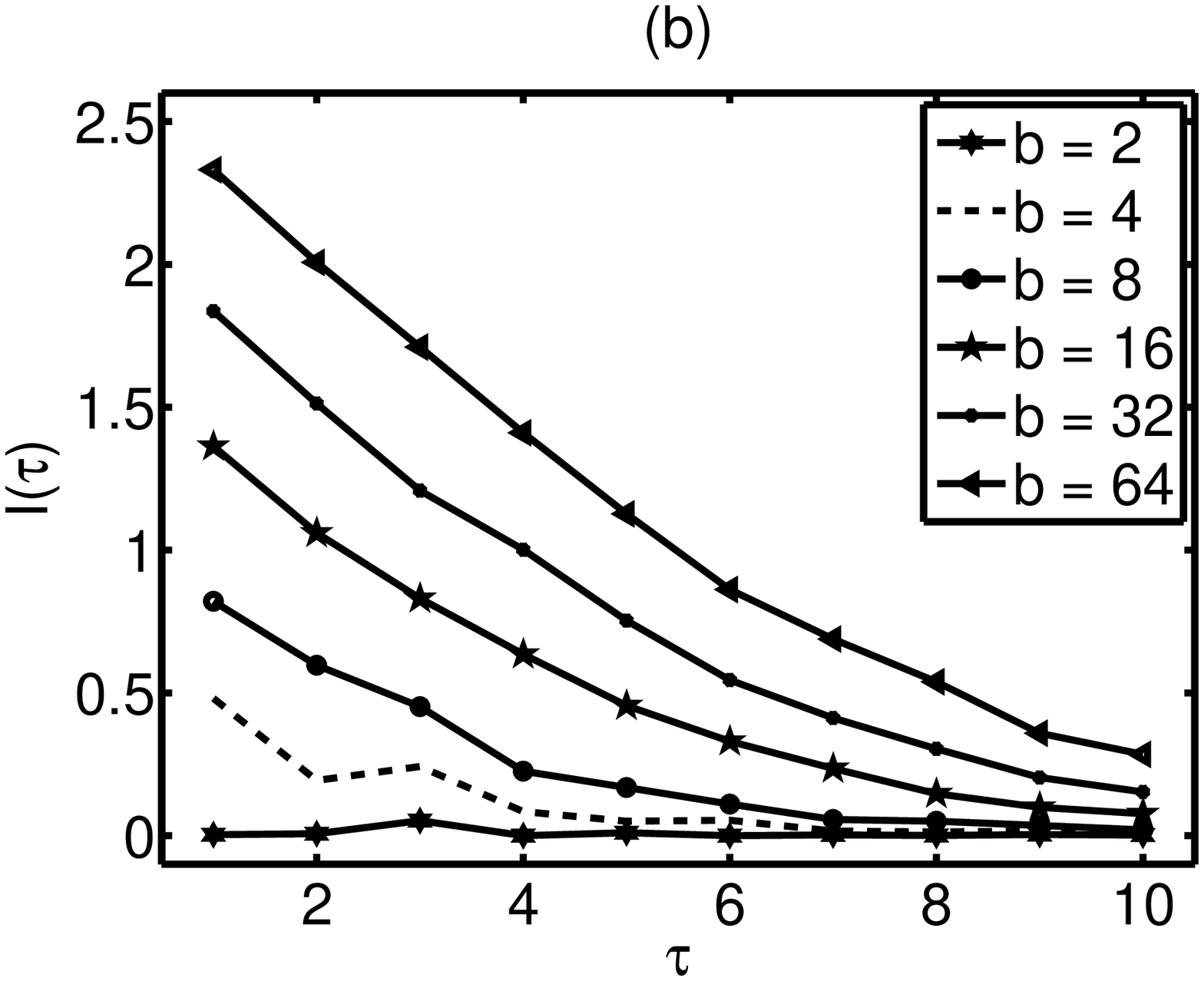}
}}
\caption{Average of ${\it I}(\tau)$ from the equidistant
histogram-based estimator for lags $\tau =1,\ldots,10$ from $1000$
simulations of the Henon map for different number of bins (as in
the legend) and for lengths $n = 1024$ in (a) and $n = 10^7$ in
(b).} \label{fig:Henonclean}
\end{figure}

%=================================
The ED estimator is heavily dependent on the selection of the
number of bins $b$. The Monte Carlo simulations showed that ${\it
I}(\tau)$ increases with $b$ even for very large time series for
all systems. Moreover, for fixed $b$ the estimator seems to be
over $n$ only for very small lags, whereas for larger $\tau$,
${\it I}(\tau)$ decreases constantly with $n$, as shown in
Fig.~\ref{fig:Henonclean} for the Henon system. For smaller values
of $b$, ${\it I}(\tau)$ is rather stable for all $n$ but gives
poor estimation close to the zero level for all lags.
%==============================================================================
% Figure 1
% ---------
%==============================================================================
The simulation on the different systems showed that the
equidistant estimator depends on $b$ and $n$, differently for
small and large lags.

\begin{figure}[h!]
 \centerline{\hbox{
\includegraphics[height=3.4cm,keepaspectratio]{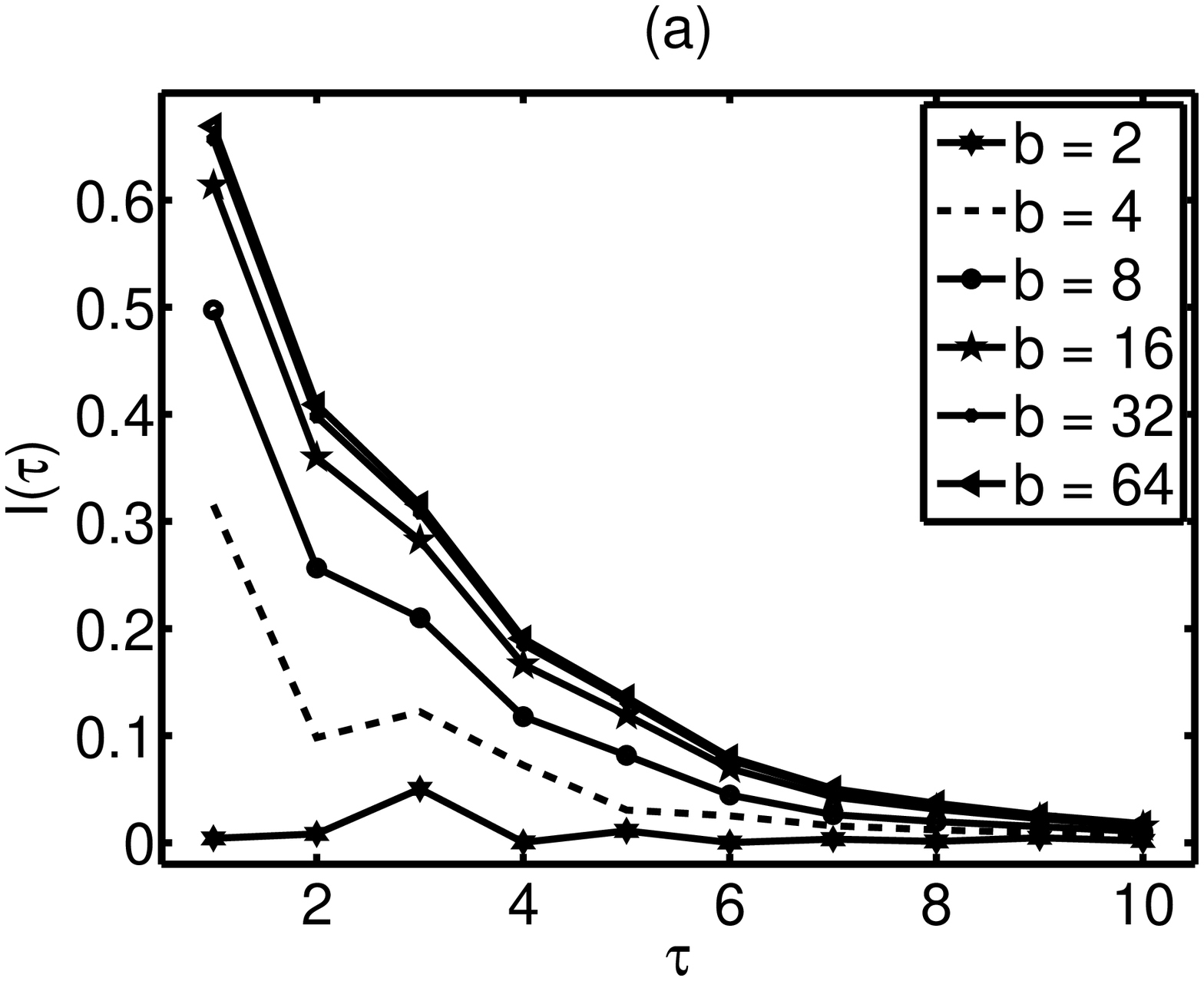}
\includegraphics[height=3.4cm,keepaspectratio]{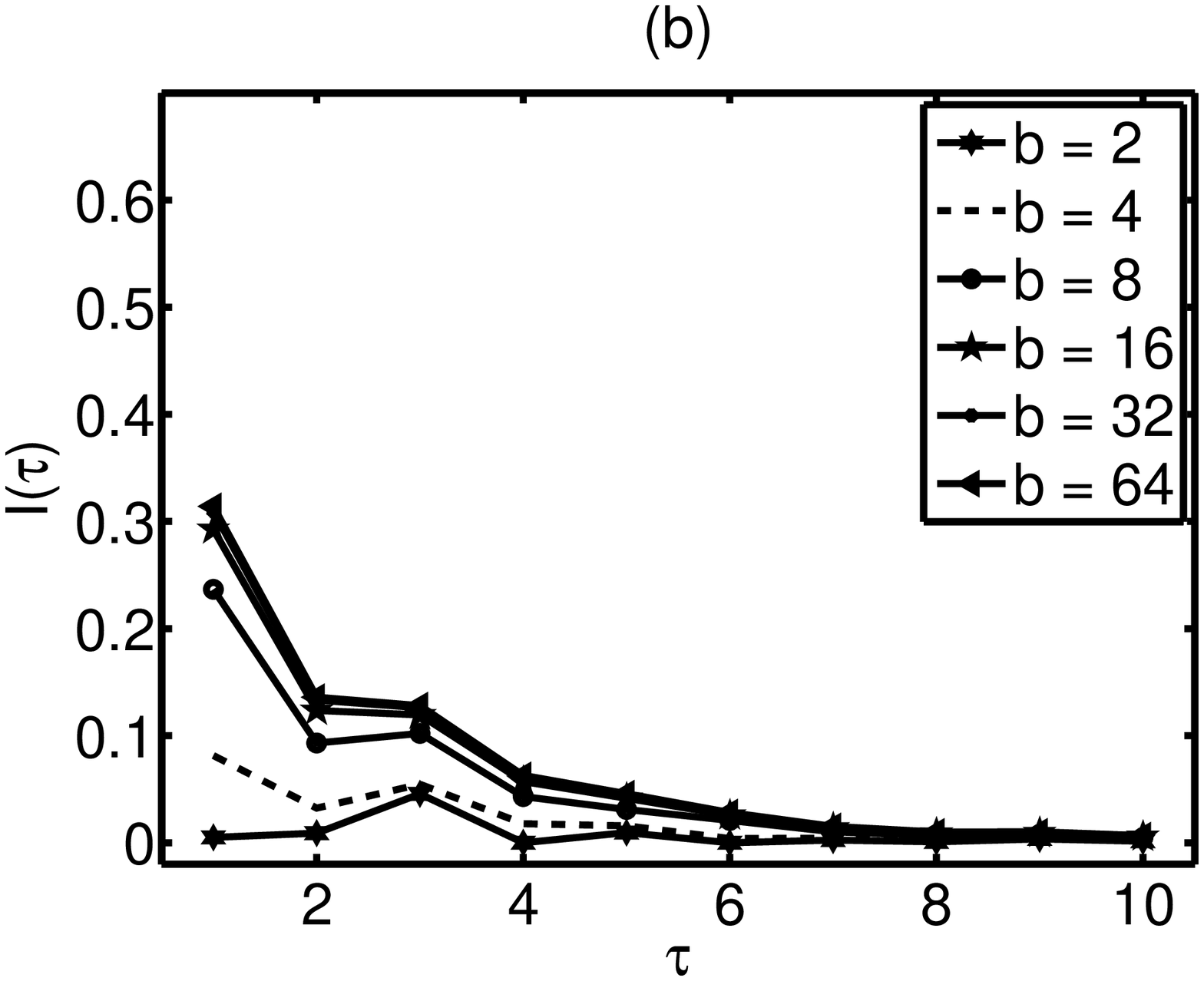}
}} \caption{${\it I}(\tau)$ from the equidistant estimator for
lags $\tau =1,\ldots,10$ from one realization of the Henon map of
length $n = 10^7$, for noise level $20 \%$ in (a) and $40 \%$ in
(b).}
 \label{fig:Henonnoisy}
\end{figure}

With the addition of noise, variations in the estimated mutual
information in terms of $n$ and $b$ become smaller. However,
this is expected as ${\it I}(\tau)$ decreases with the increase
of noise level (see Fig.\ref{fig:Henonnoisy}).

%==============================================================================
% Figure 2
% --------
%==============================================================================

The inclusion of stronger noise component masks the
deterministic structure and thus levels the estimate of mutual
information towards zero. On the other hand, the benefit of
noise in terms of stable estimation, is that ${\it I}(\tau)$
gets less dependent on $b$, e.g. in Fig.\ref{fig:Henonnoisy}
the estimate converges when $b>16$ for noise level at $20\%$
and $b>8$ for noise level at $40\%$.

\begin{figure}[h!]
\centerline{\hbox{
\includegraphics[height=3.4cm,keepaspectratio]{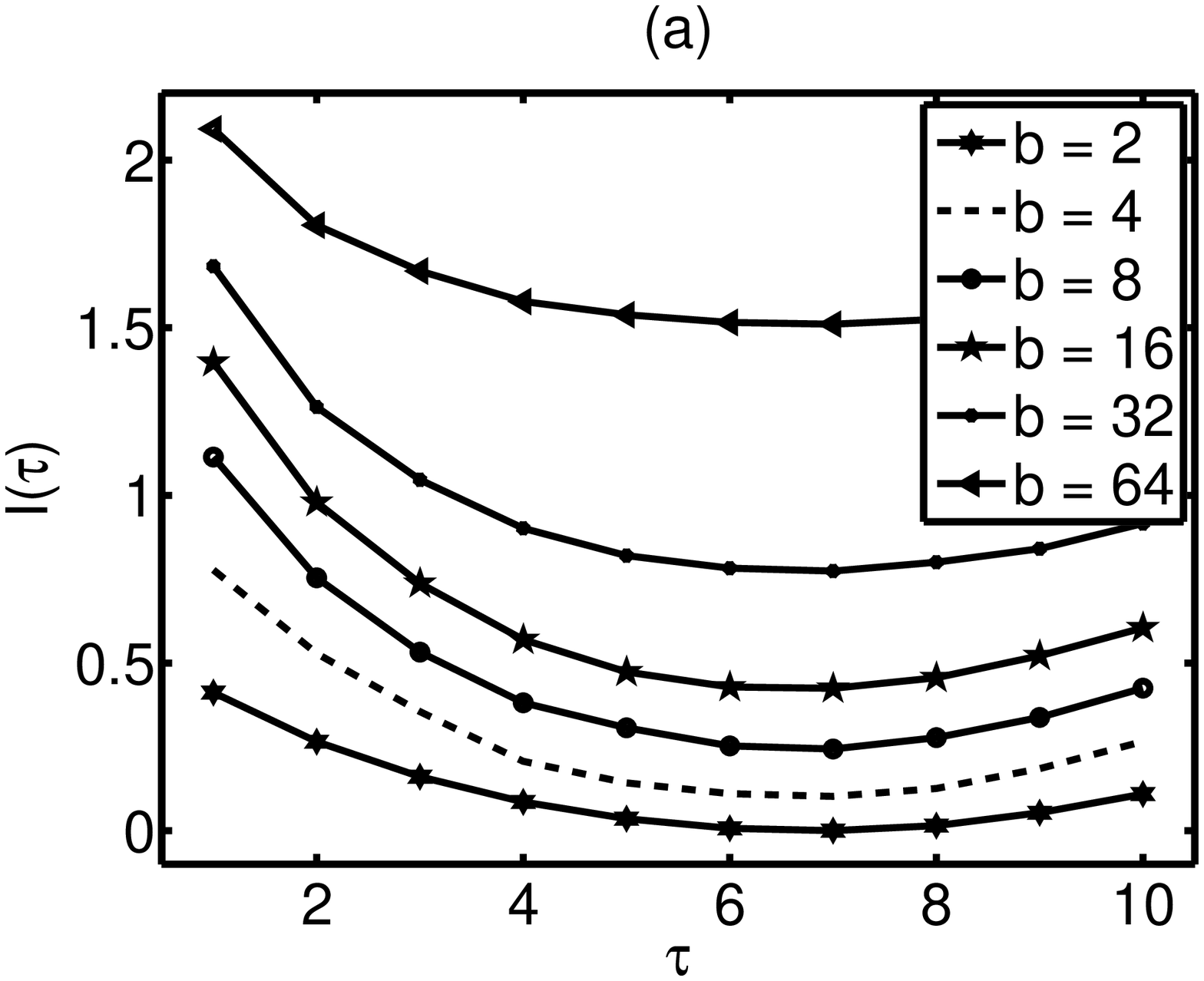}
\includegraphics[height=3.4cm,keepaspectratio]{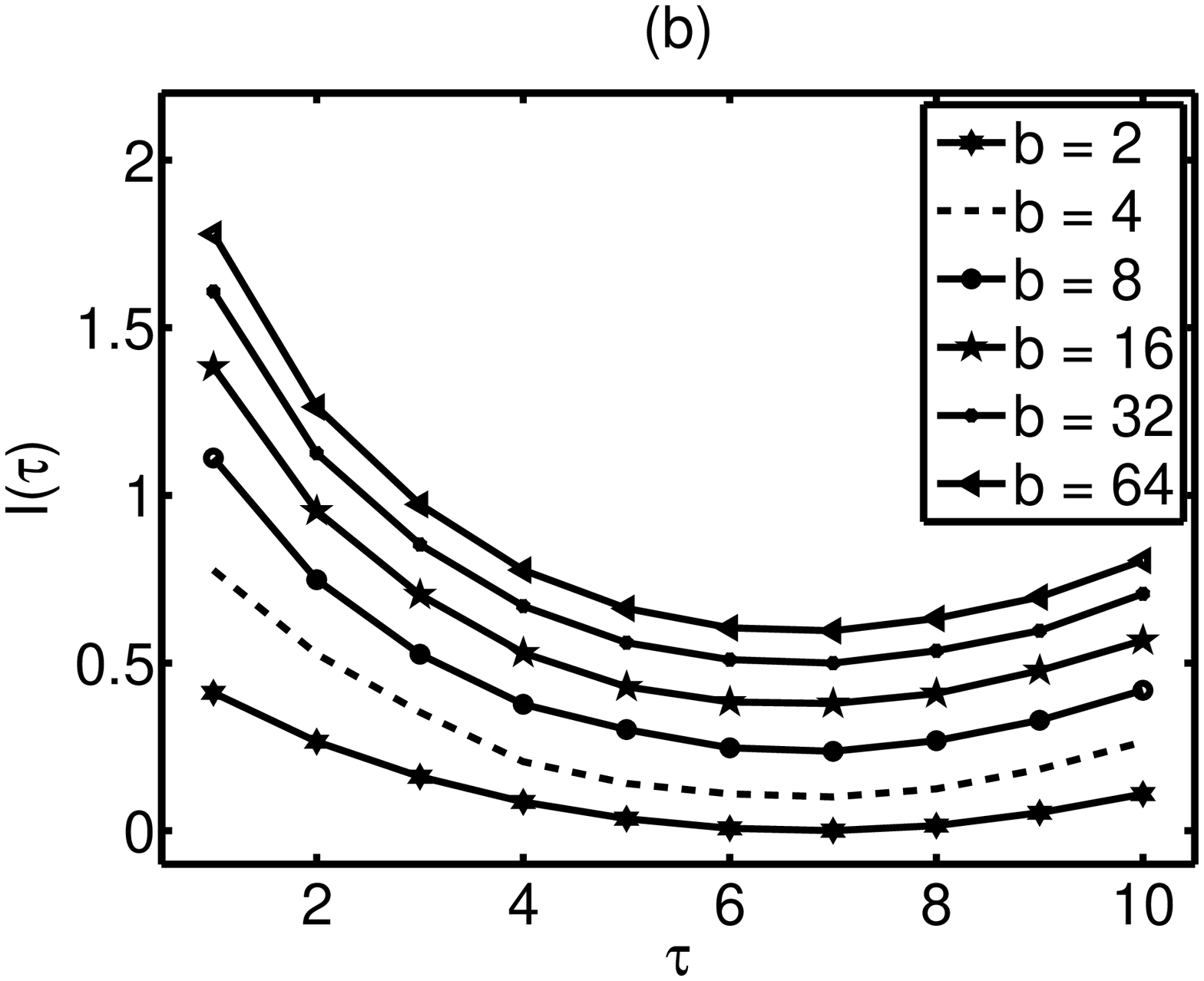}
}}
\caption{Average of ${\it I}(\tau)$ from the equidistant
estimator for lags $\tau =1,\ldots,10$ from $1000$ realizations of
the Mackey Glass system (sampling time 17) with $\Delta = 17$ of
length $n = 1024$ in (a) and ${\it I}(\tau)$ from one realization
of length $n = 10^6$ in (b).} \label{fig:MGclean}
\end{figure}

For Mackey-Glass system, the ${\it I}(\tau)$ is computed for a
range of lags that include the lag of first minimum of mutual
information in order to assess also the accuracy of the
estimator in detecting this particular lag. We observed that
although ${\it I}(\tau)$ increases with $b$, the lag of the
first minimum of mutual information does not vary with $b$.
Moreover, the estimate of this lag is rather stable with $n$,
as shown in Fig.\ref{fig:MGclean} for noise-free data and
$\Delta = 17$.
%==============================================================================
% Figure 3
% --------
%==============================================================================

\subsection{Equiprobable estimator}
\begin{figure}[h!]
 \centerline{\hbox{
\includegraphics[height=3.4cm,keepaspectratio]{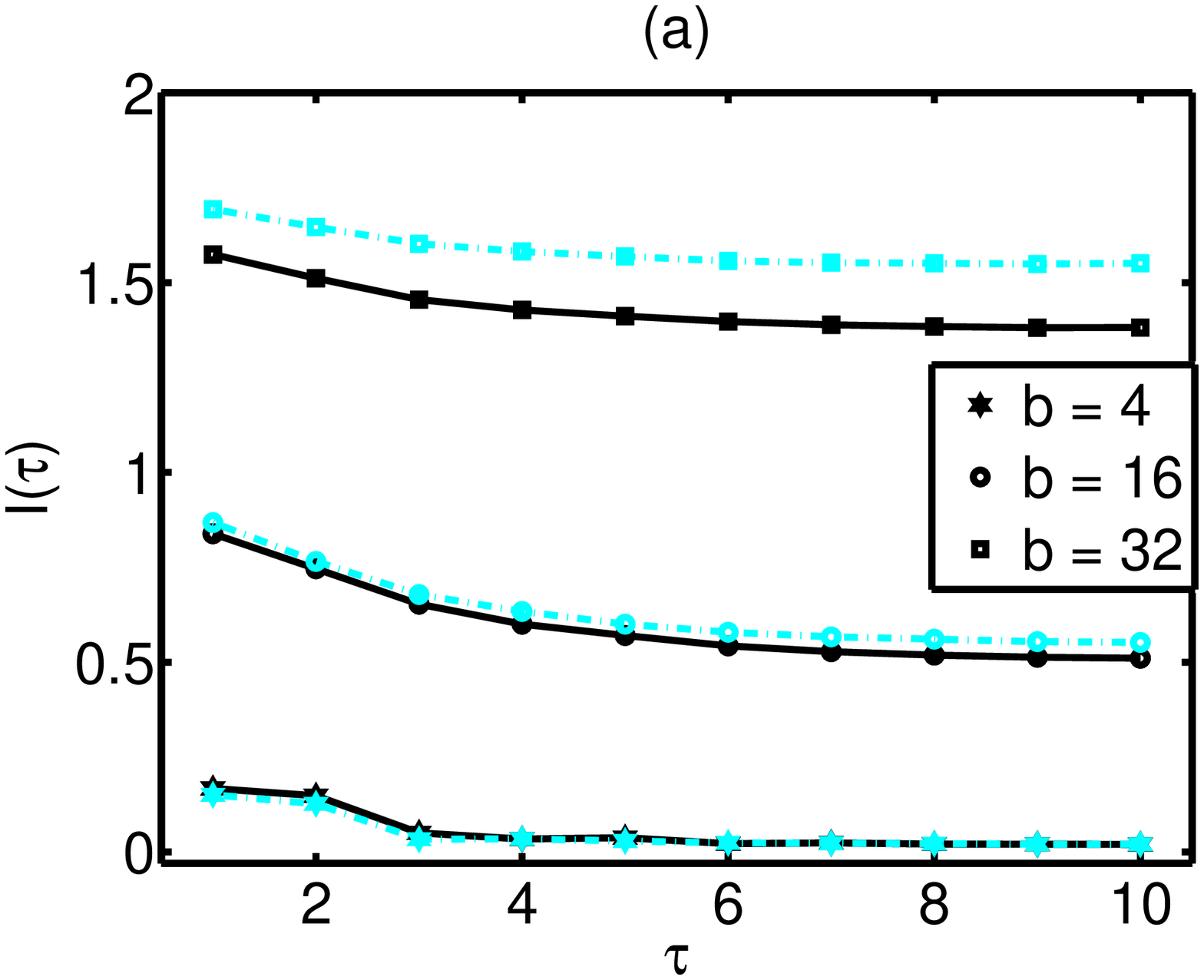}
\includegraphics[height=3.4cm,keepaspectratio]{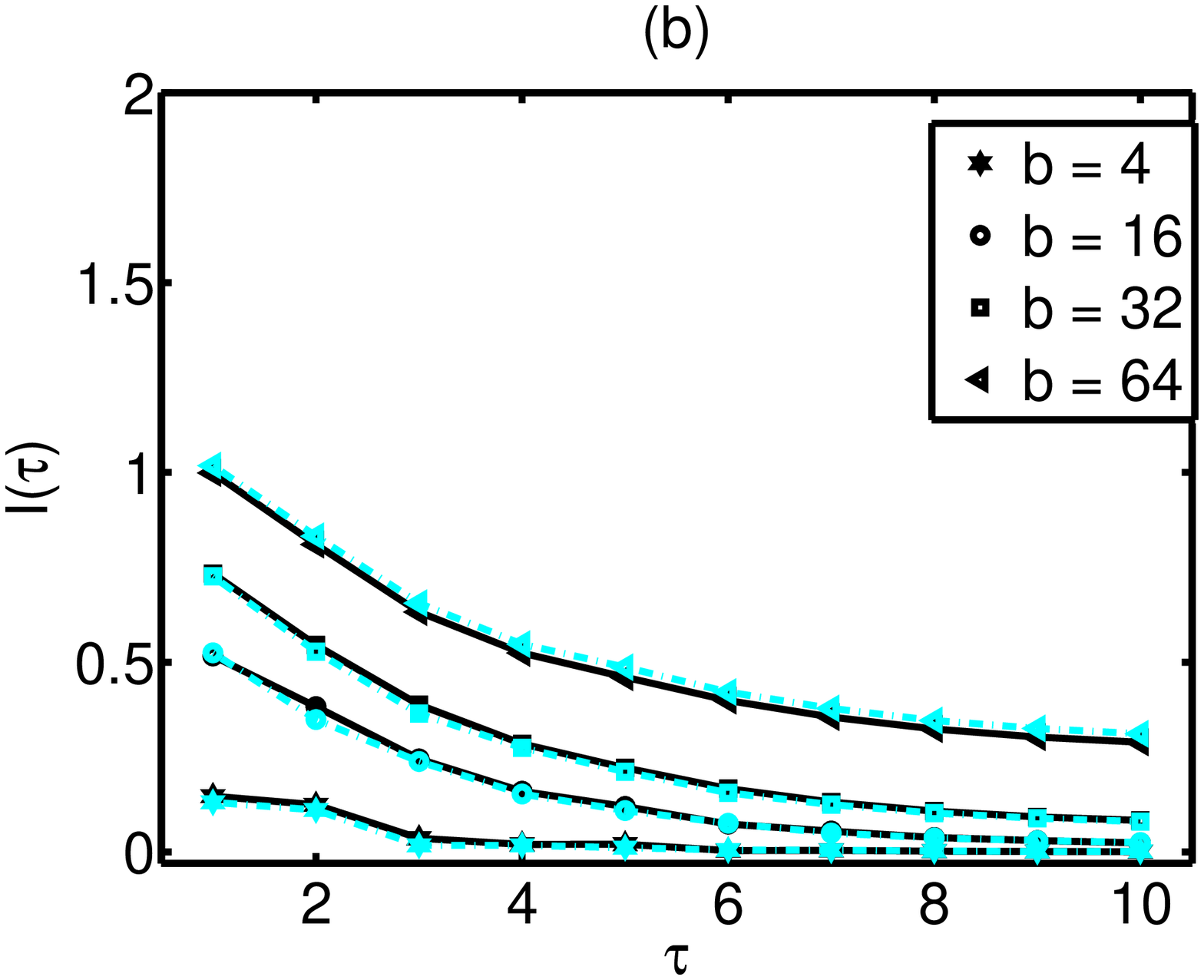}
}}
\caption{Average of ${\it I}(\tau)$ from the equidistant
estimator (black lines) and the equiprobable estimator (gray
lines, cyan in the color online) for lags $\tau=1,\ldots,10$ from
$1000$ realizations of the Ikeda map for different number of bins
(as in the legend), for lengths $n = 256$ in (a) and $n = 8192$ in
(b).}
 \label{fig:Ikedaclean}
\end{figure}

%=================================
The EP estimator has the same form of dependence on $b$, $n$ and
$\tau$ as the ED estimator. Moreover, the estimated ${\it
I}(\tau)$ values from the two estimators may vary for small $n$
but converge with $n$, as shown for the noise-free data from the
Ikeda system in Fig.\ref{fig:Ikedaclean}.

%==============================================================================
% Figure 4
% --------
%==============================================================================

Other works seem to agree with the result that estimators using a
fixed partition heavily depend on the selection of $b$
\cite{Bonnlander94,Cellucci05,Trappenberg06}. Given that the
chaotic systems have significant mutual information at least for
very small lags and that the ED and EP estimators get close to
zero for small number of bins, we conclude that a small $b$ is a
good choice for these estimators only for independent time series.
(We observed the same for linear stochastic systems in a different
study.) To the contrary, for chaotic systems a refined partition
explores in more detail the fine structure of the distribution
(marginal and joint) and gives better estimate of the true mutual
information. We would expect that in the limit of fine partition,
where the expression of the discretized version of mutual
information in \eqref{eq:def2} converges to the true mutual
information in \eqref{eq:def1}, ${\it I}(\tau)$ would converge as
well, but this would require an infinite amount of data. Thus a
good choice of bin width should balance a large $b$ with the
limited size $n$ of the available data in order to maintain a good
estimation of the probabilities in \eqref{eq:def2}, in particular
at regions with low data densities that may carry valuable
information for the system dynamics. For a fixed $b$, the ED and
EP estimators are rather stable with respect to $n$, i.e. they are
consistent estimators of the mutual information as defined in
\eqref{eq:def2} for the specific partition determined by $b$.

%=============================================
\subsection{Adaptive histogram-based estimator}
%=============================================

\begin{figure}[h!]
 \centerline{\hbox{
\includegraphics[height=3.4cm,keepaspectratio]{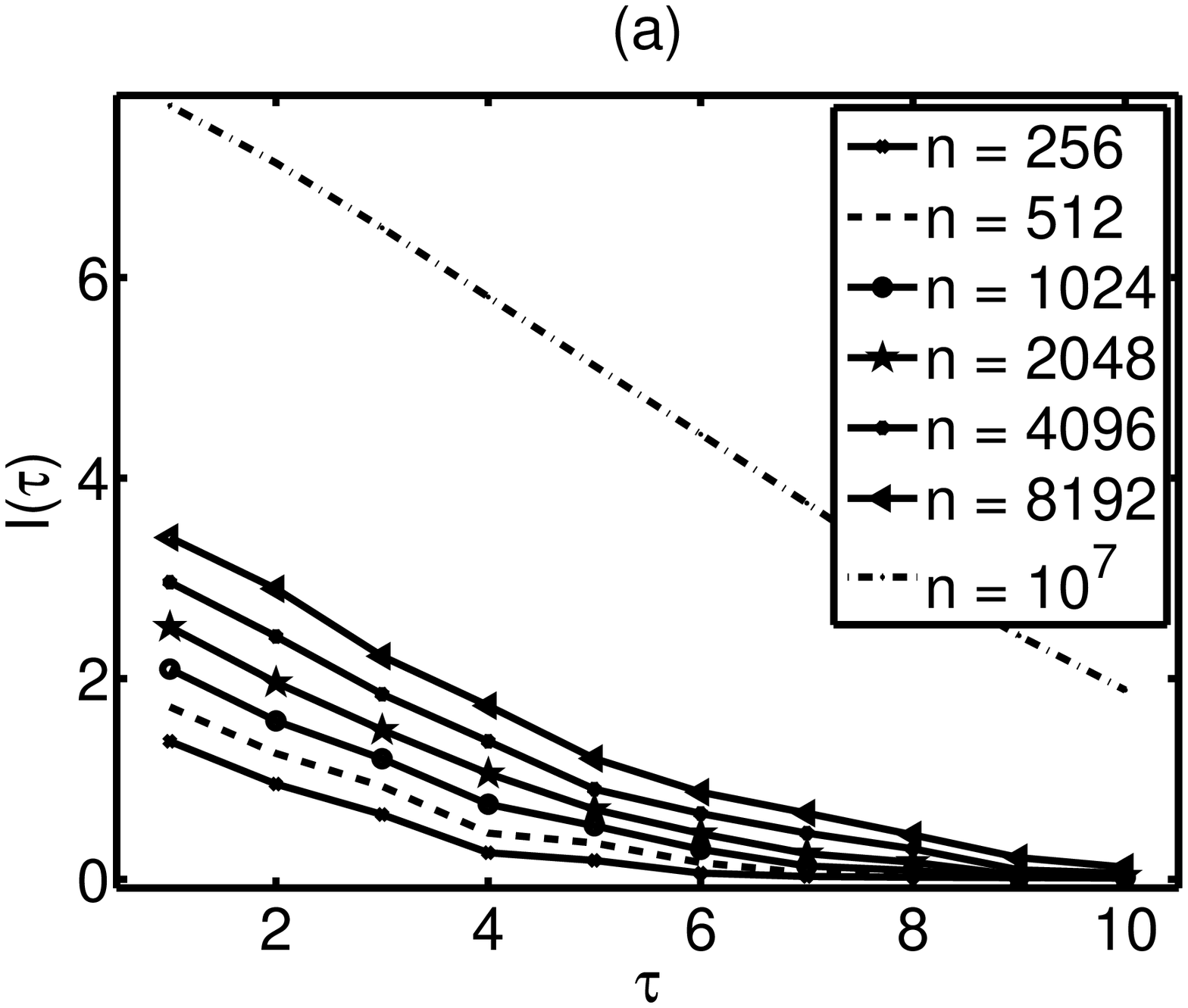}
\includegraphics[height=3.4cm,keepaspectratio]{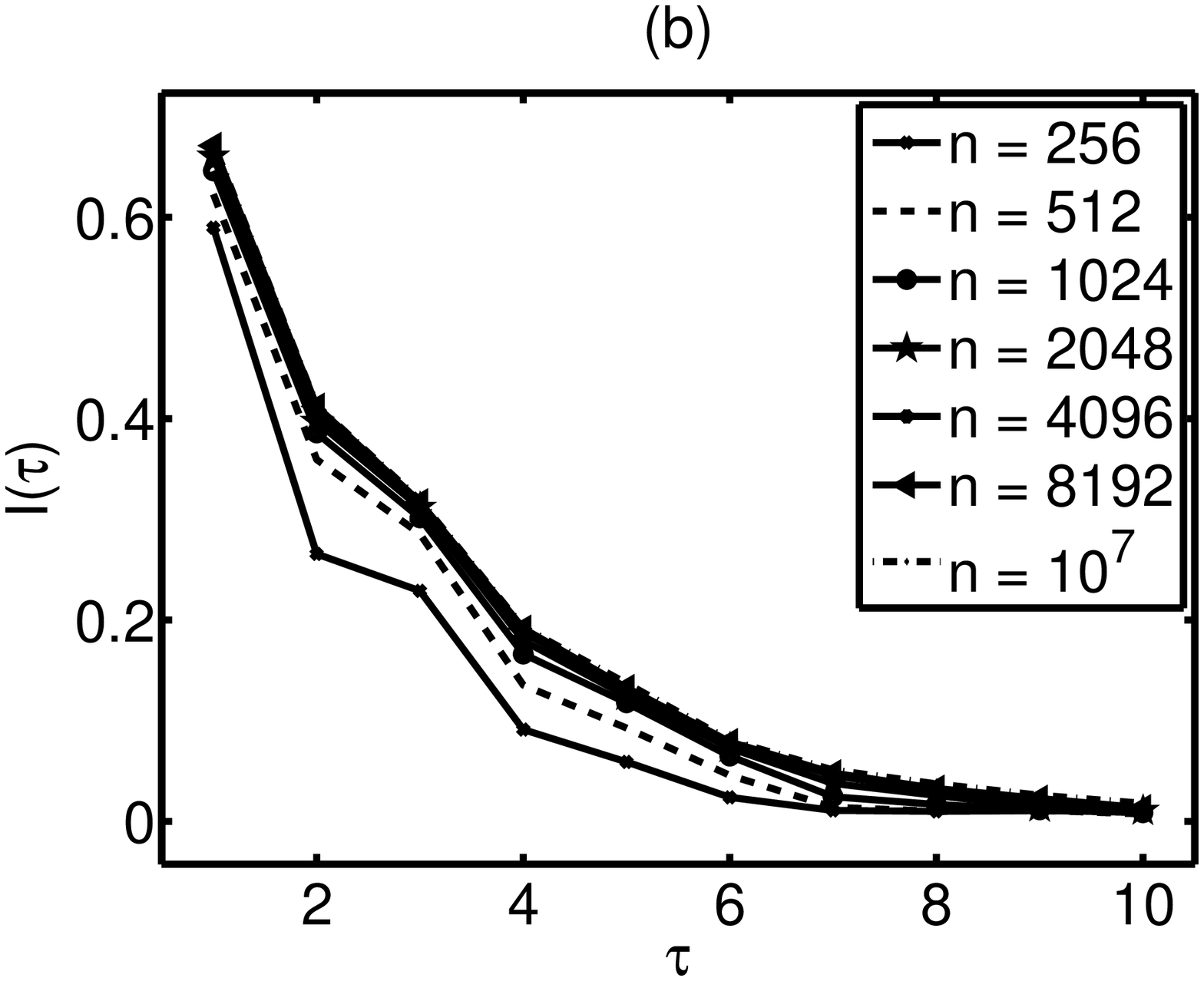}
}}
\caption{Average of ${\it I}(\tau)$ from the adaptive
histogram-based estimator for lags $\tau =1,\ldots,10$ from $1000$
realizations of the Henon map for different lengths $n$ (as in the
legend), for noise-free data in (a) and noisy data at $20\%$ noise
level in (b).}
 \label{fig:Henonadaptive}
\end{figure}

The AD estimator of Darbellay and Vajda is apparently independent
of a parameter for the partitioning. However, it has a direct
dependence on $n$, which determines the roughness of the
partitioning in a somehow automatic way. In the abundance of data,
the AD estimator reaches a very fine partition that satisfies the
independence condition in each cell, so that the total number of
cells is very large (analogously to a fixed-partition with a
respectively large $b$). Thus the increase of $n$ implies a finer
partition and explains the increase of ${\it I}(\tau)$ from the
adaptive estimator with $n$, as shown for the Henon map in
Fig.\ref{fig:Henonadaptive}a. Note that the dependence of AD on
$n$ is not comparable to that of the fixed-bin estimators because
it involves a change of partitioning with $n$. In this sense, AD
is not consistent with respect to $n$. However, in the presence of
noise, the effect of $n$ on the adaptive estimator decreases with
the noise level (see Fig.\ref{fig:Henonadaptive}b).

%==============================================================================
% Figure5
% =======
%==============================================================================

AD is considered to be one of the most precise and efficient
algorithms for estimating the mutual information that converges
fast to the true mutual information, basically for certain
distribution and Gaussian processes where this is analytically
given \cite{Kraskov04,Trappenberg06}. However, in the case of
nonlinear systems the estimator does not seem to converge with
$n$, unless the fine partition is limited by the presence of
noise.

%=============================================
\subsection{$k$-nearest neighbor estimator}
%=============================================

\begin{figure}[h!]
\centerline{\hbox{
\includegraphics[height=3.4cm,keepaspectratio]{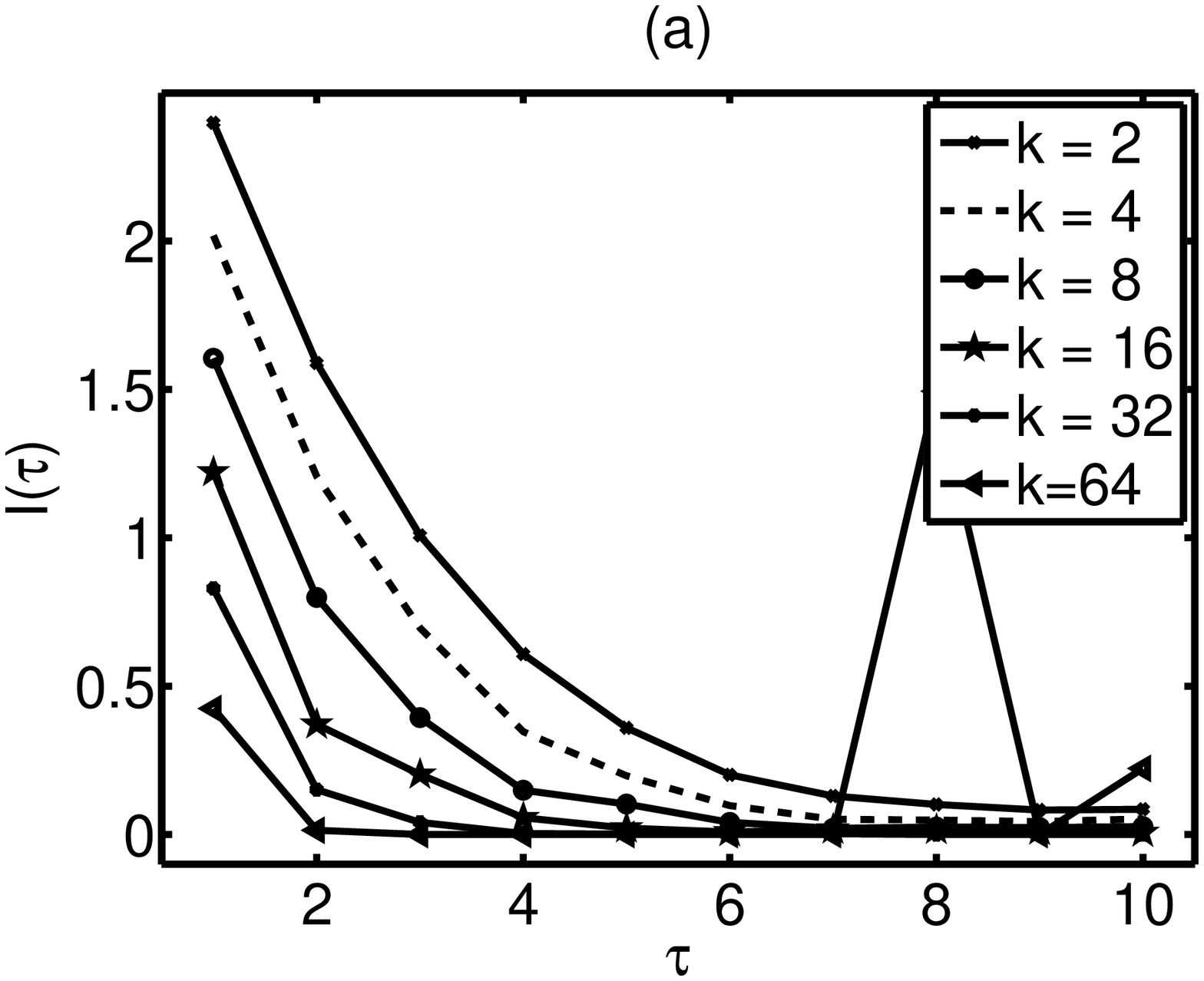}
\includegraphics[height=3.4cm,keepaspectratio]{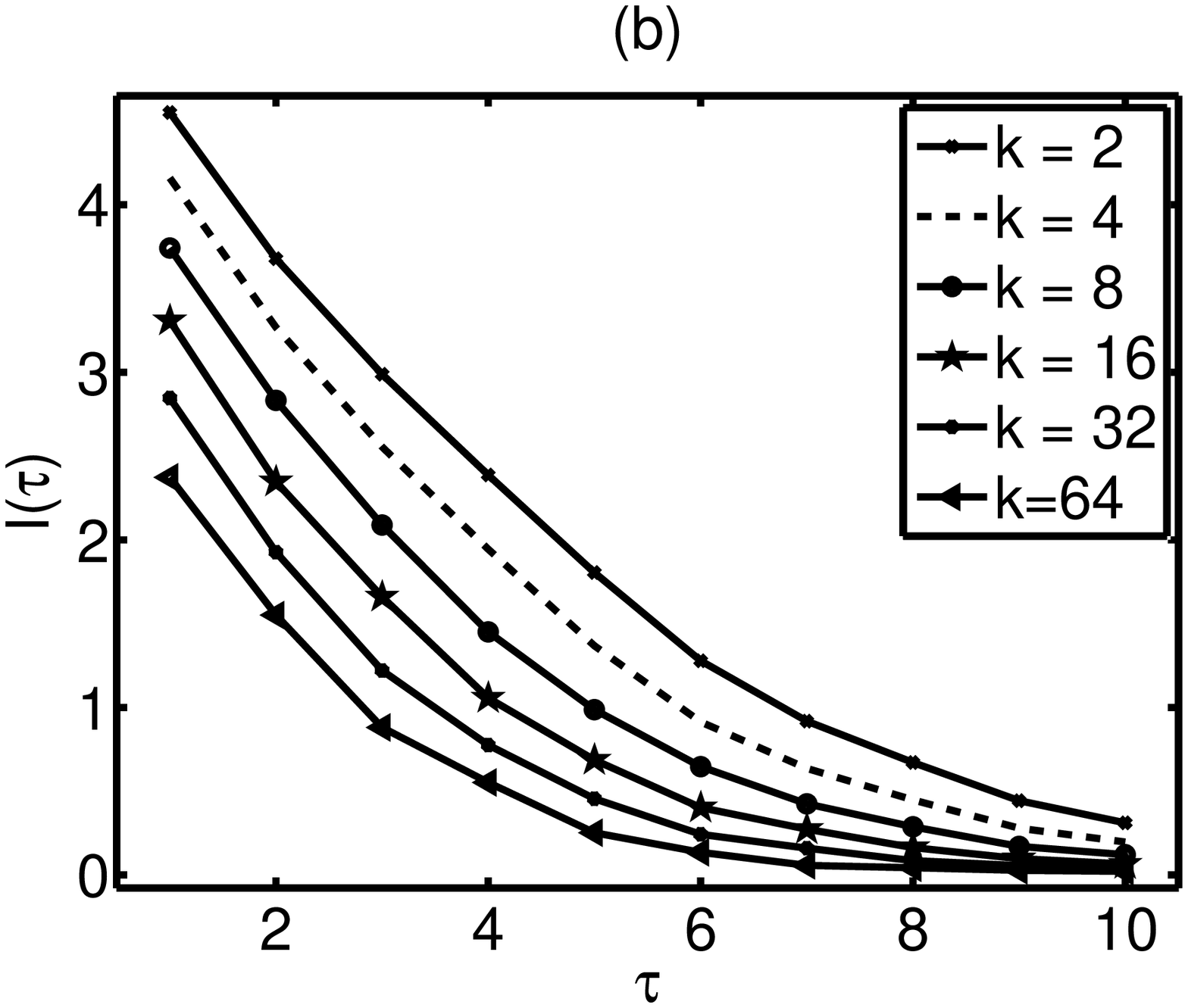}
}}
\centerline{\hbox{
\includegraphics[height=3.4cm,keepaspectratio]{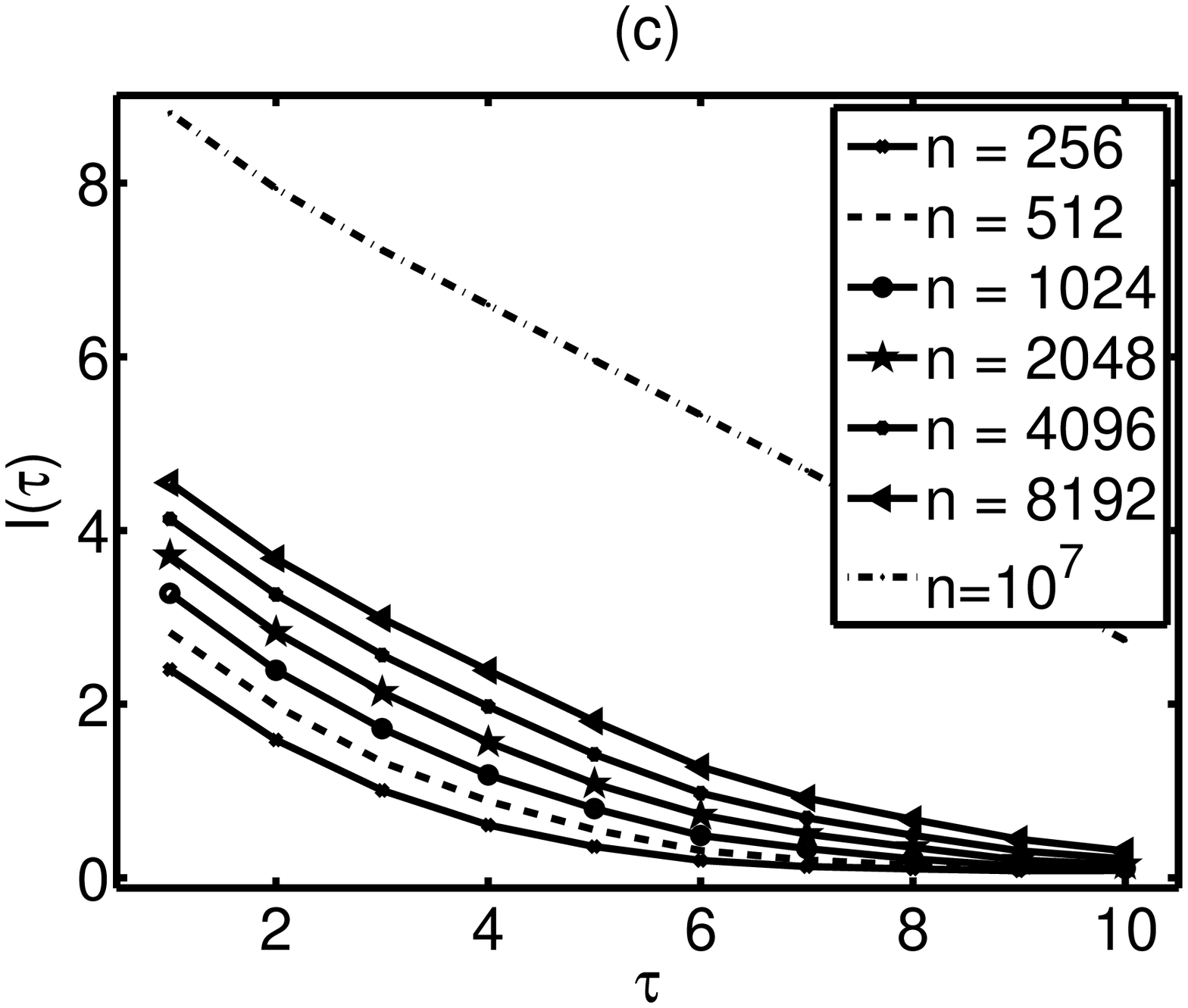}
\includegraphics[height=3.4cm,keepaspectratio]{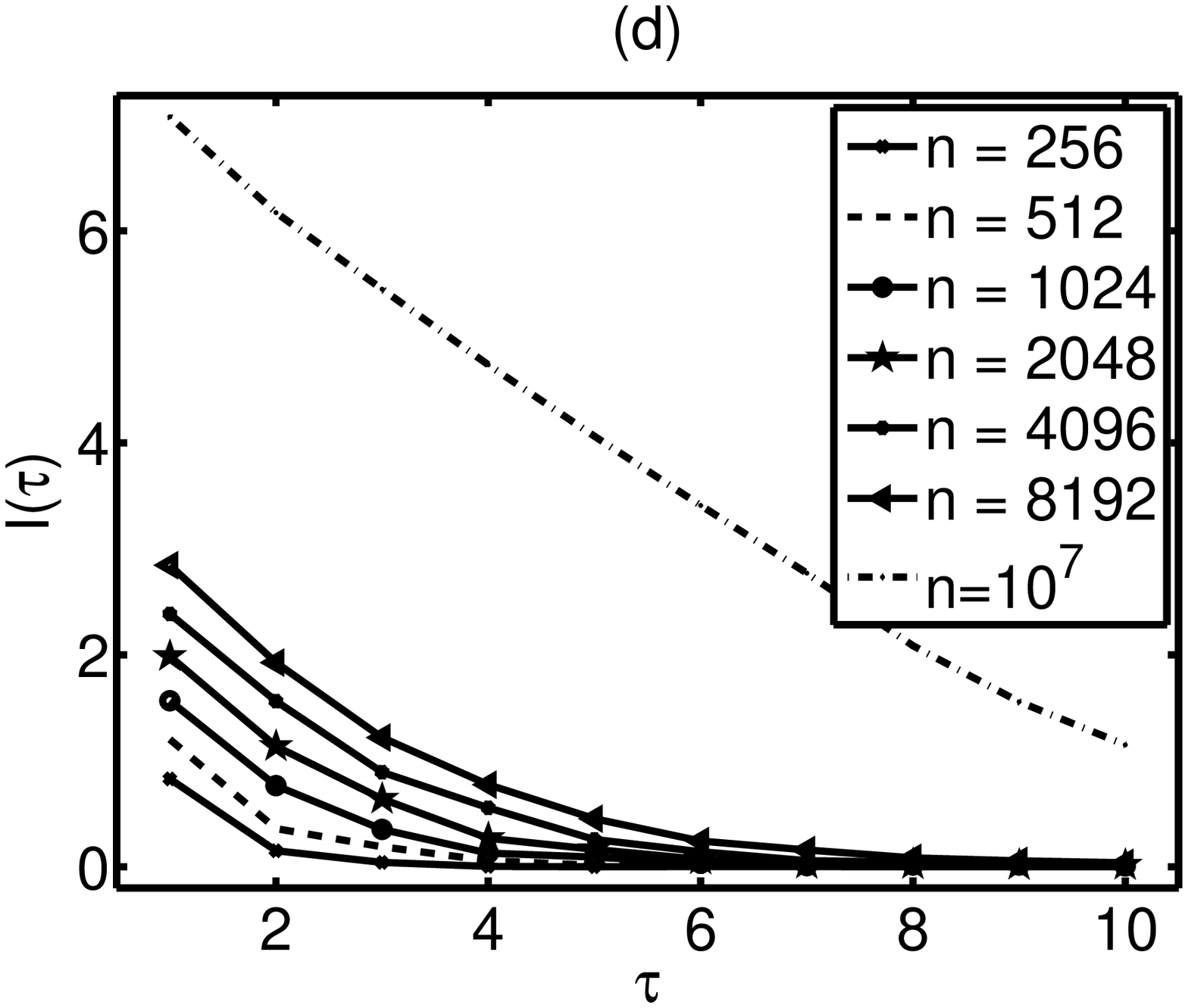}
}}
\caption{Average of ${\it I}(\tau)$ from the $k$-nearest
neighbor estimator for lags $\tau =1,\ldots,10$ from $1000$
realizations of the Henon map. In (a) and (b) the results are
displayed for different $k$ (as in the legend) and for $n = 256$
and $n = 8192$, respectively. In (c) and (d) the results are
displayed for different $n$ (as in the legend) and for $k = 2$ and
$k = 32$, respectively.} \label{fig:Henonneighbors}
\end{figure}

The parameter of the number of nearest neighbors $k$ in this
estimator determines the roughness of approximation of the density
functions in \eqref{eq:def1}, which corresponds to the roughness
of the partitioning in \eqref{eq:def2}. Thus for a fixed $n$, as
${\it I}(\tau)$ from fixed-bin estimators increase with $b$ (finer
partition), the ${\it I}(\tau)$ from KNN increases when $k$
decreases, as shown in Fig.\ref{fig:Henonneighbors}a and b.
%==============================================================================
% Figure6
% =======
%==============================================================================
This dependence persists also for very large $n$, whereas for a
small time series a large $k$ gives a poor estimation of the
densities and consequently of ${\it I}(\tau)$ (see
Fig.\ref{fig:Henonneighbors}a, particularly when $k=64$ and
$n=256$).

We also observed that ${\it I}(\tau)$ increase with $n$ for a
fixed $k$ (see Fig.\ref{fig:Henonneighbors}c and d). Assuming a
fixed parameter ($k$ and $b$) the effect of $n$ on KNN is larger
than on the fixed-bin estimators and similar to the effect of $n$
on the adaptive histogram-based estimator.

\begin{figure}[h!]
 \centerline{\hbox{
\includegraphics[height=3.4cm,keepaspectratio]{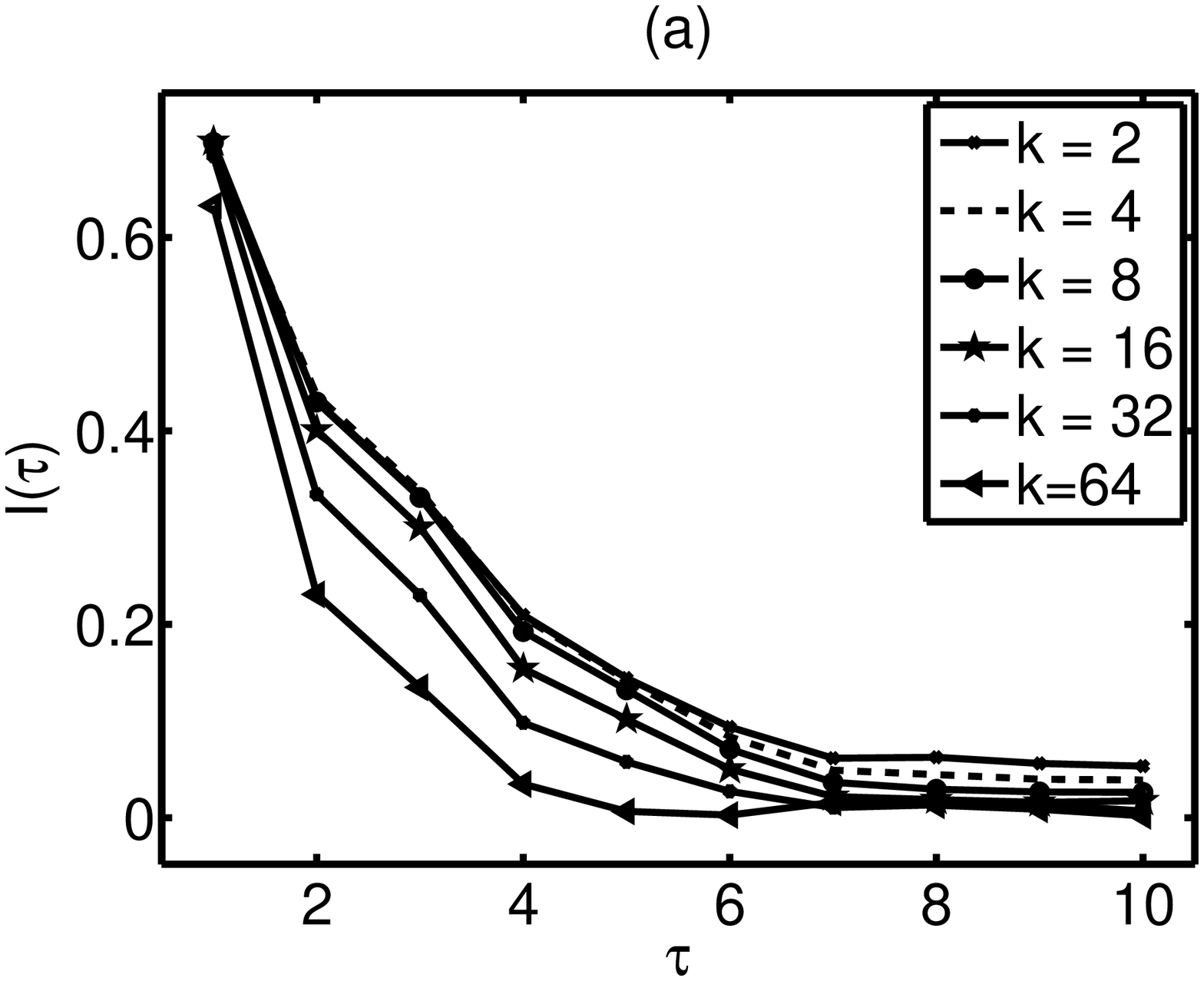}
\includegraphics[height=3.4cm,keepaspectratio]{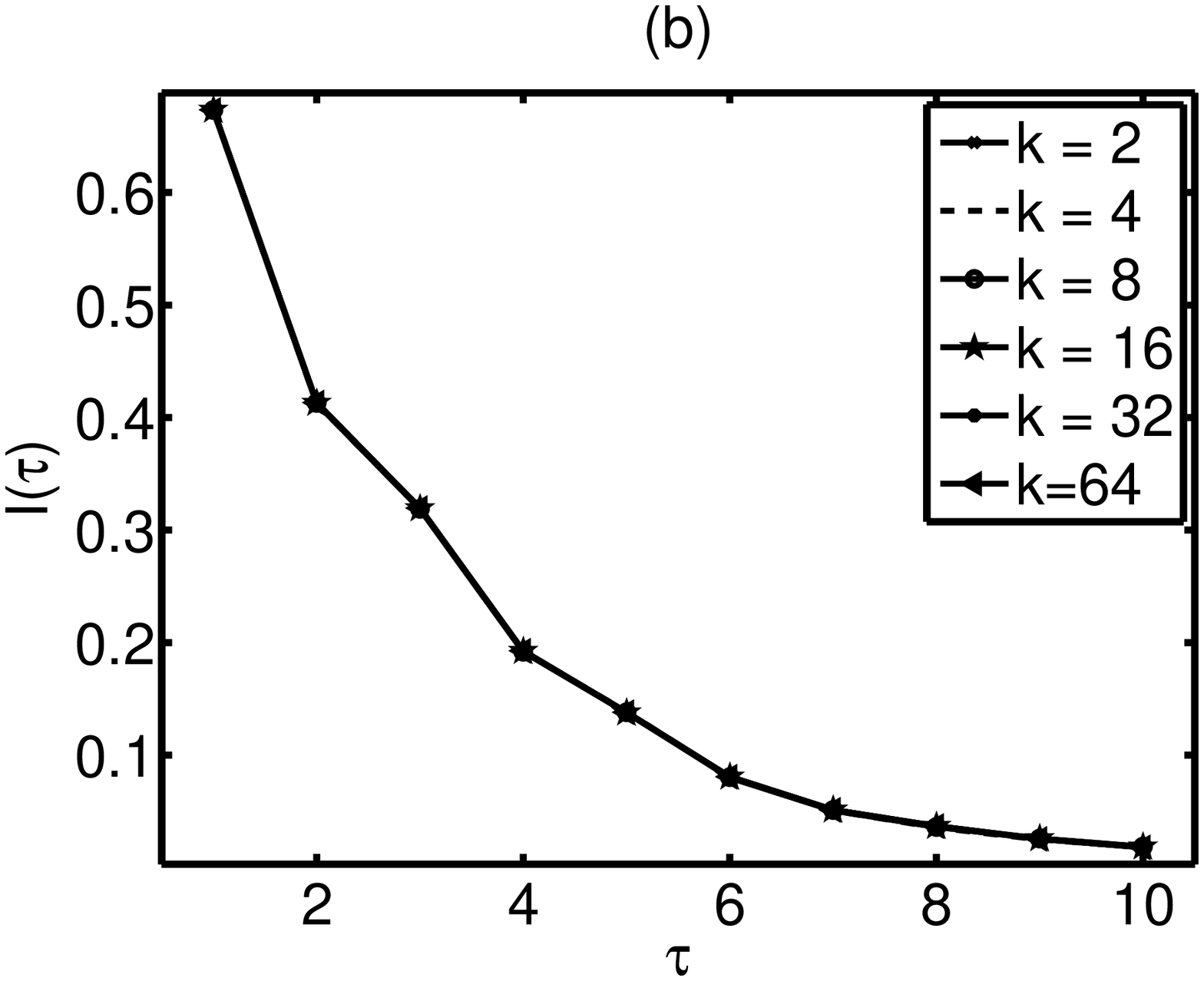}
}} \caption{Average of ${\it I}(\tau)$ from the $k$-nearest
neighbor estimator for lags $\tau =1,\ldots,10$ from $1000$
realizations of the Henon map with $20 \%$ noise of length $n =
1024$ in (a) and $I(\tau)$ from one realization of length $n =
10^7$ in (b).} \label{fig:Henonnoisyneighbors}
\end{figure}

In agreement with the histogram-based estimators, KNN decreases
with the noise level (see Fig.\ref{fig:Henonnoisyneighbors}).
%==============================================================================
% Figure7
% =======
%==============================================================================
However in the presence of noise, the dependence of the KNN on $k$
is less than the dependence of the ED and EP on $b$ (e.g. for the
case of Henon map with $n=10^7$ and $20\%$ noise level compare
Fig.\ref{fig:Henonnoisyneighbors}b to Fig.\ref{fig:Henonnoisy}a).
This results is in agreement with Nicolaou et al.
\cite{Nicolaou05} that found independence of this estimator on $k$
for EEG data.

An upper limit for $k$ is set by $n$ whereas the lower limit $k=1$
corresponds to the finest partition we could get for a
histogram-based estimator. Thus the restriction to small $k$
proposed by Kraskov et al. \cite{Kraskov04} and used in other
simulation studies \cite{Kreuz07,Khan07}, corresponds to fine
partitions. Thus for chaotic systems that require fine
partitioning, a fair comparison of the KNN for these $k$ values to
ED and EP would require a very large $b$. To this respect, this
simulation comparison is restricted to comparable small $b$ (up to
64) due to computational limitations.

%==============================
\subsection{Kernel estimator}
%===============================
Among different kernel functions used in the literature for
density estimation, and for mutual information estimation in
particular, the common practice is to use the Gaussian kernel in
conjunction with the "Gaussian" bandwidth of Silverman
\cite{Moon95} or multiplies of it
\cite{Harrold01,Steuer02,Khan07}. There are some other criteria
for bandwidth selection that we have included in this study (see
Table~\ref{tab:bandwidth}). Moreover, we test the KE estimator of
mutual information also for a range of bandwidths as for the other
estimators.

\begin{figure}[h!]
 \centerline{\hbox{
\includegraphics[height=3.4cm,keepaspectratio]{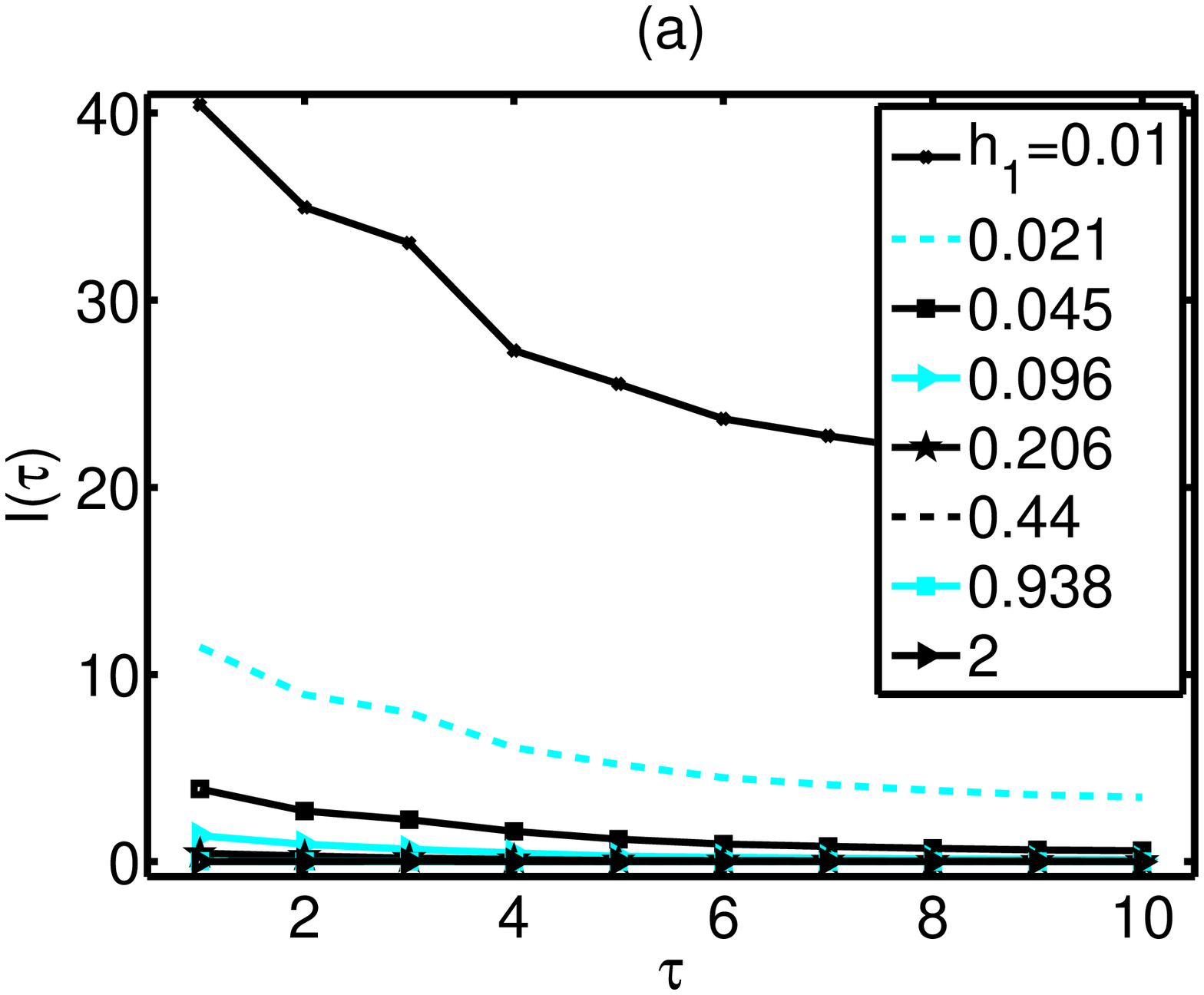}
\includegraphics[height=3.4cm,keepaspectratio]{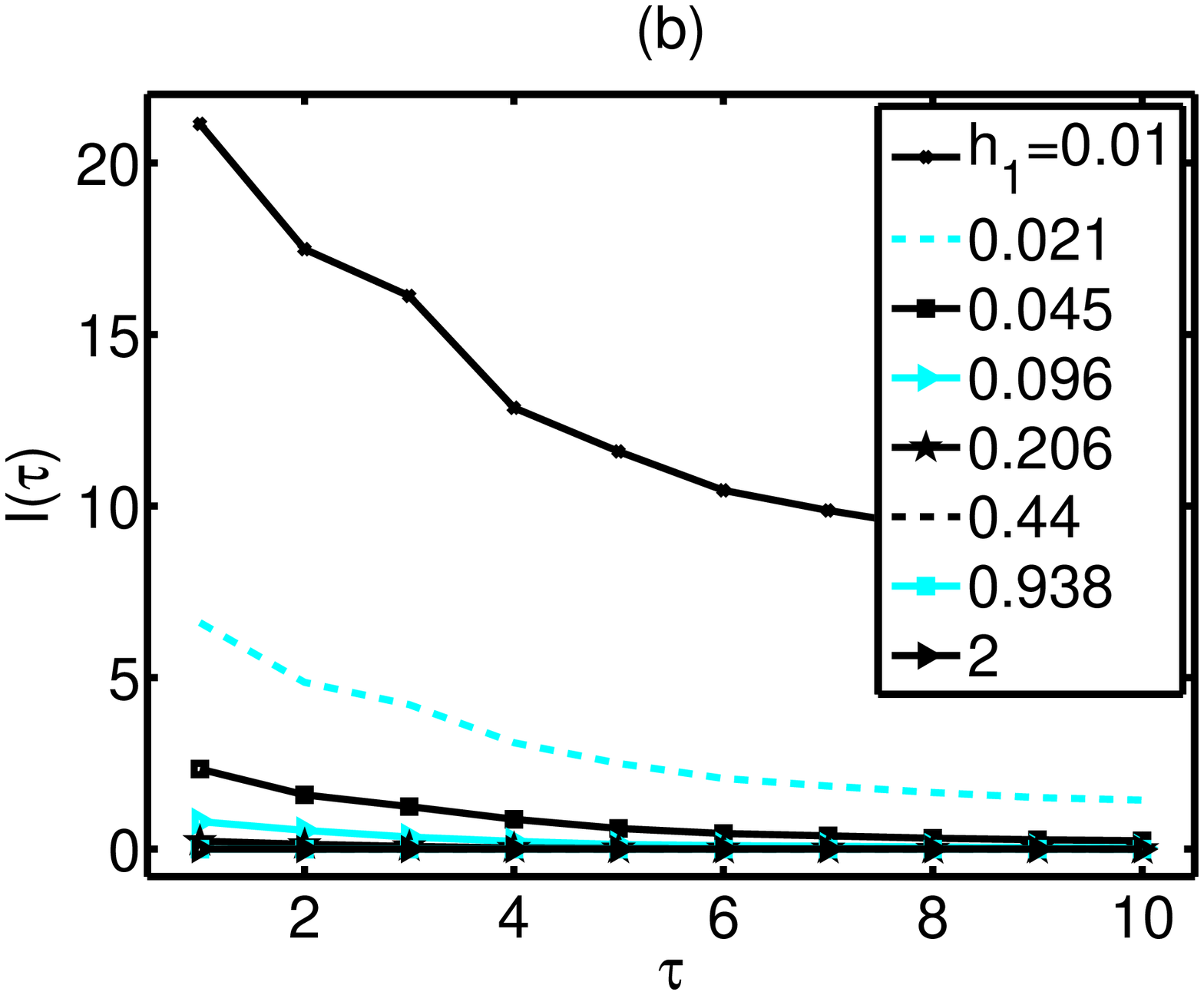}
}}
 \centerline{\hbox{
\includegraphics[height=3.4cm,keepaspectratio]{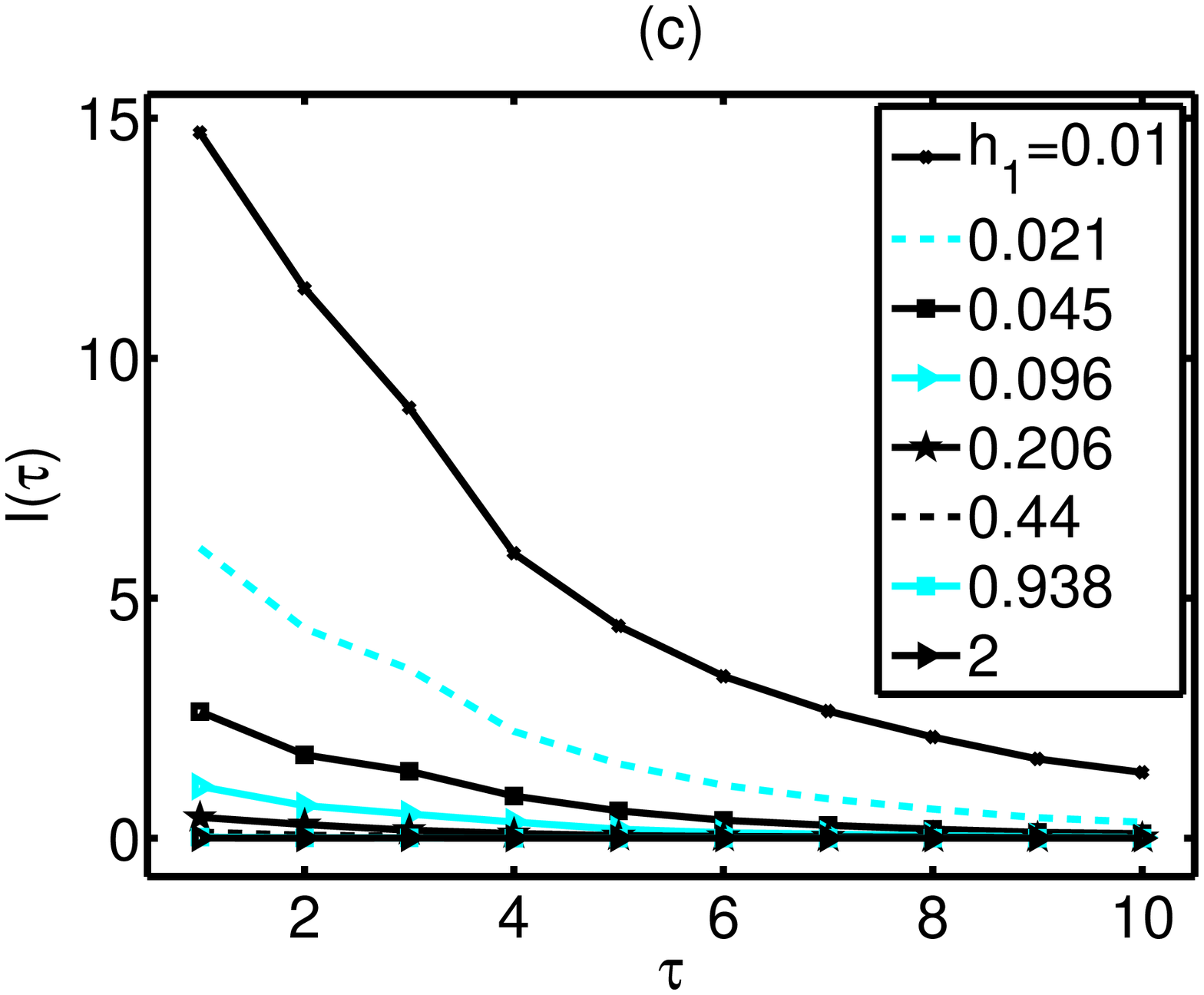}
\includegraphics[height=3.4cm,keepaspectratio]{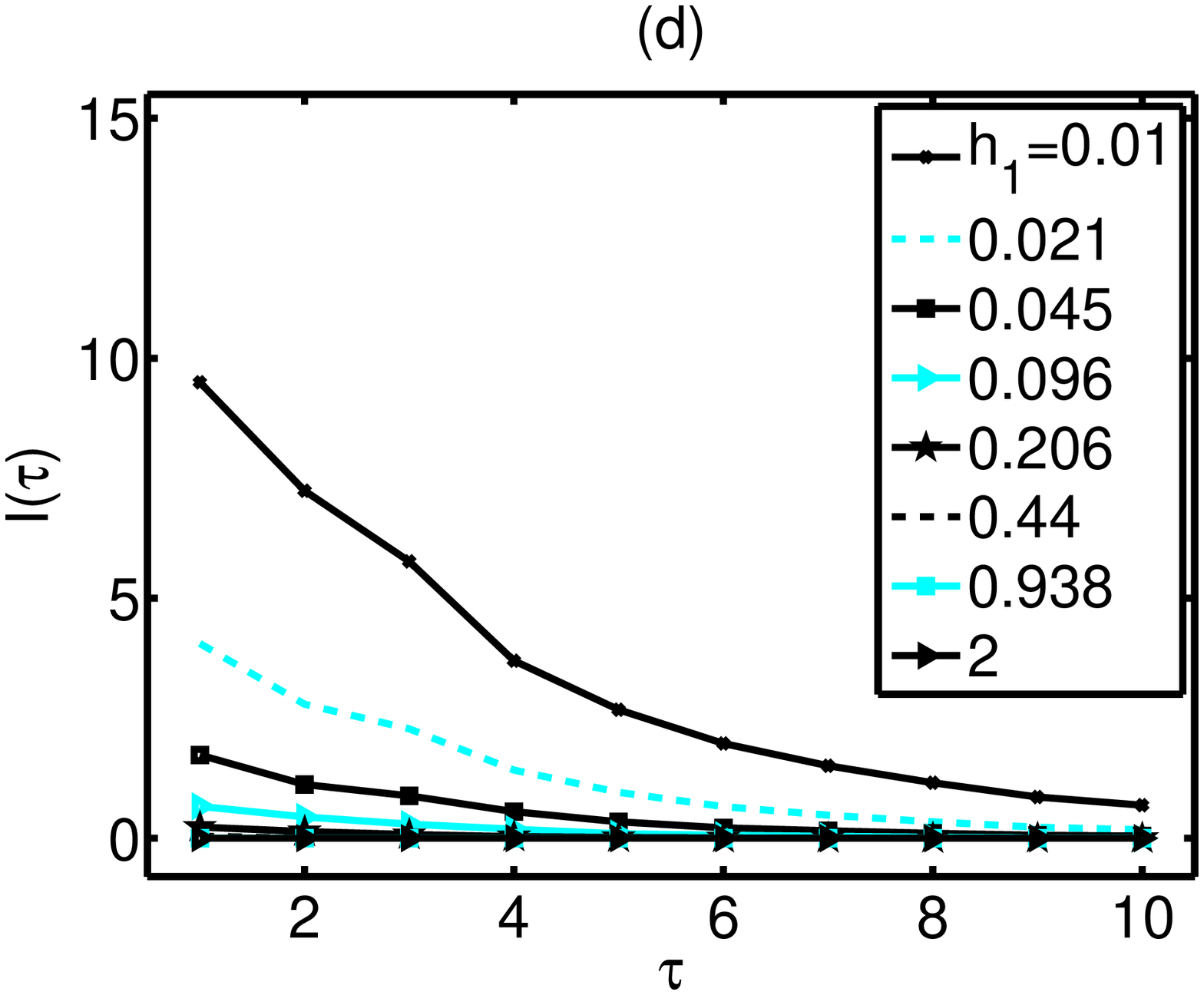}
}}
\caption{Average of ${\it I}(\tau)$ from the kernel estimator
for lags $\tau =1,\ldots,10$ from $1000$ realizations of the Henon
map for a range of different values of bandwidth $h_1$ (as in the
legend). (a) $n=512$ and $h_2=h_1$, (b) $n=512$ and
$h_2=\sqrt{2}h_1$, (c) $n=8192$ and $h_2=h_1$, (d) $n=8192$ and
$h_2=\sqrt{2}h_1$}
\label{fig:Henonkernelbandsweep}
\end{figure}

First we investigate the inter-dependence in the selection of
$h_1$ and $h_2$ across the selected range of bandwidths. As shown
in Fig.\ref{fig:Henonkernelbandsweep} for the Henon map, selecting
$h_2=\sqrt{2}h_1$ instead of $h_2=h_1$ simply decreases the value
of $I(\tau)$, without alerting the form of the dependence of
$I(\tau)$ on $\tau$ and this occurs independently of $n$.
% ==============================================================================
% Figure8
% =======
%==============================================================================
So, the roughness of the partitioning is determined by $h_1$, i.e.
smaller $h_1$ implies finer partitions or small neighborhoods with
respect to KNN. We note however that $\it{I}(\tau)$ from KEreaches
very small values for as large $h_1$ as 2 and very large values
for as small $h_1$ as $0.01$. Note that such extremely large
values of ${\it I}(\tau)$ do not occur by any other estimator.

\begin{figure}[h!]
\centerline{\hbox{
\includegraphics[height=3.4cm,keepaspectratio]{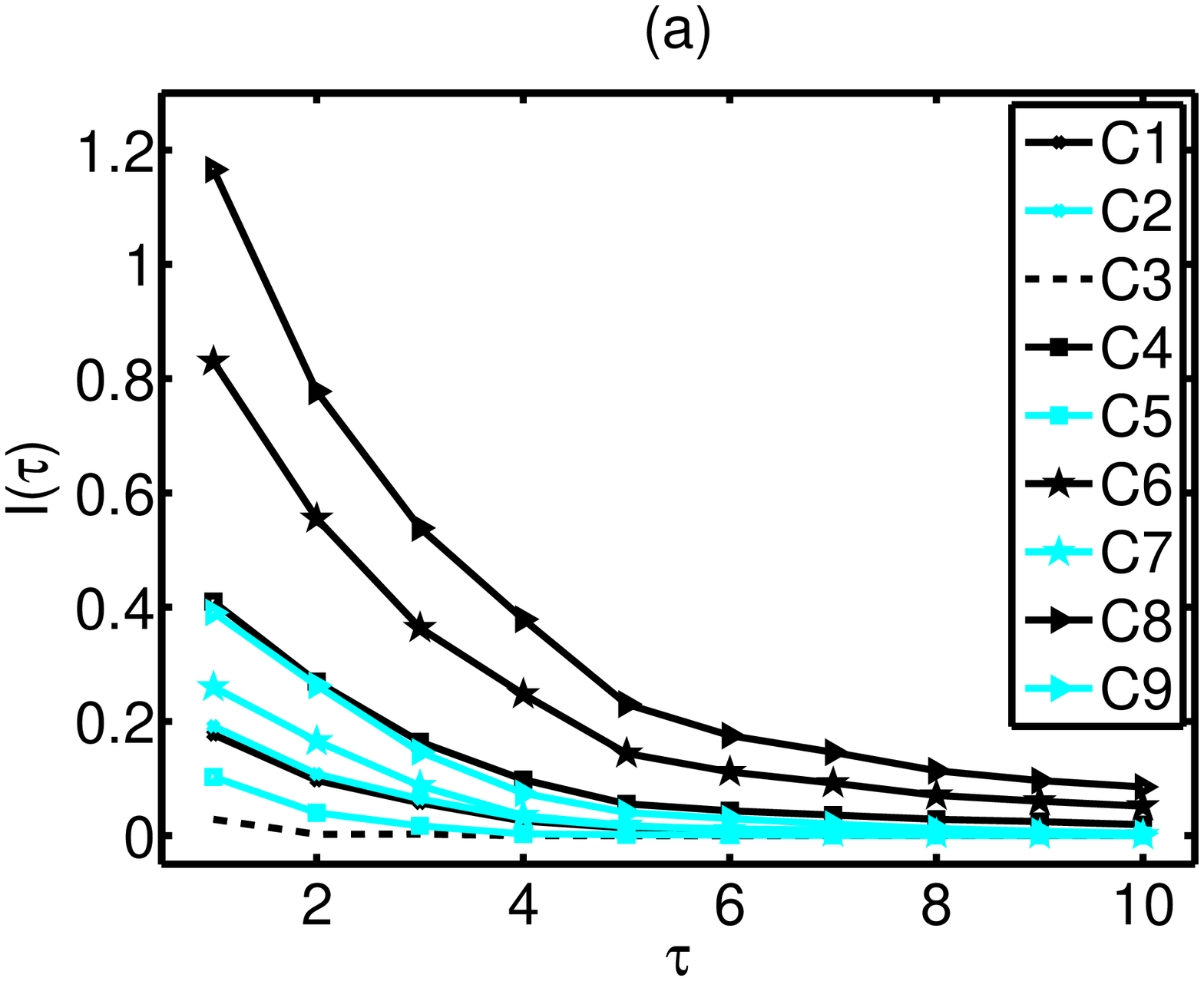}
\includegraphics[height=3.4cm,keepaspectratio]{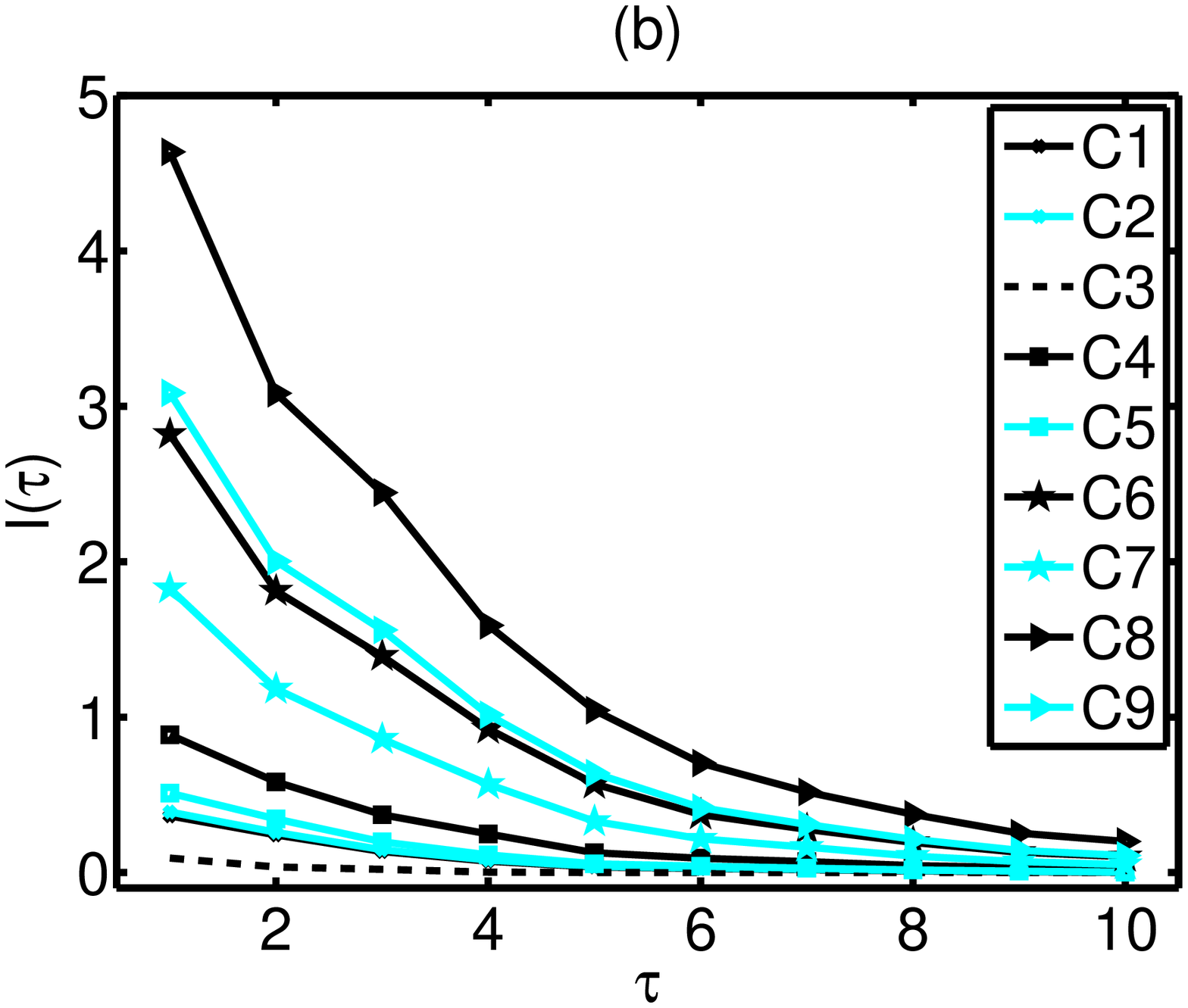}
}}
 \centerline{\hbox{
\includegraphics[height=3.4cm,keepaspectratio]{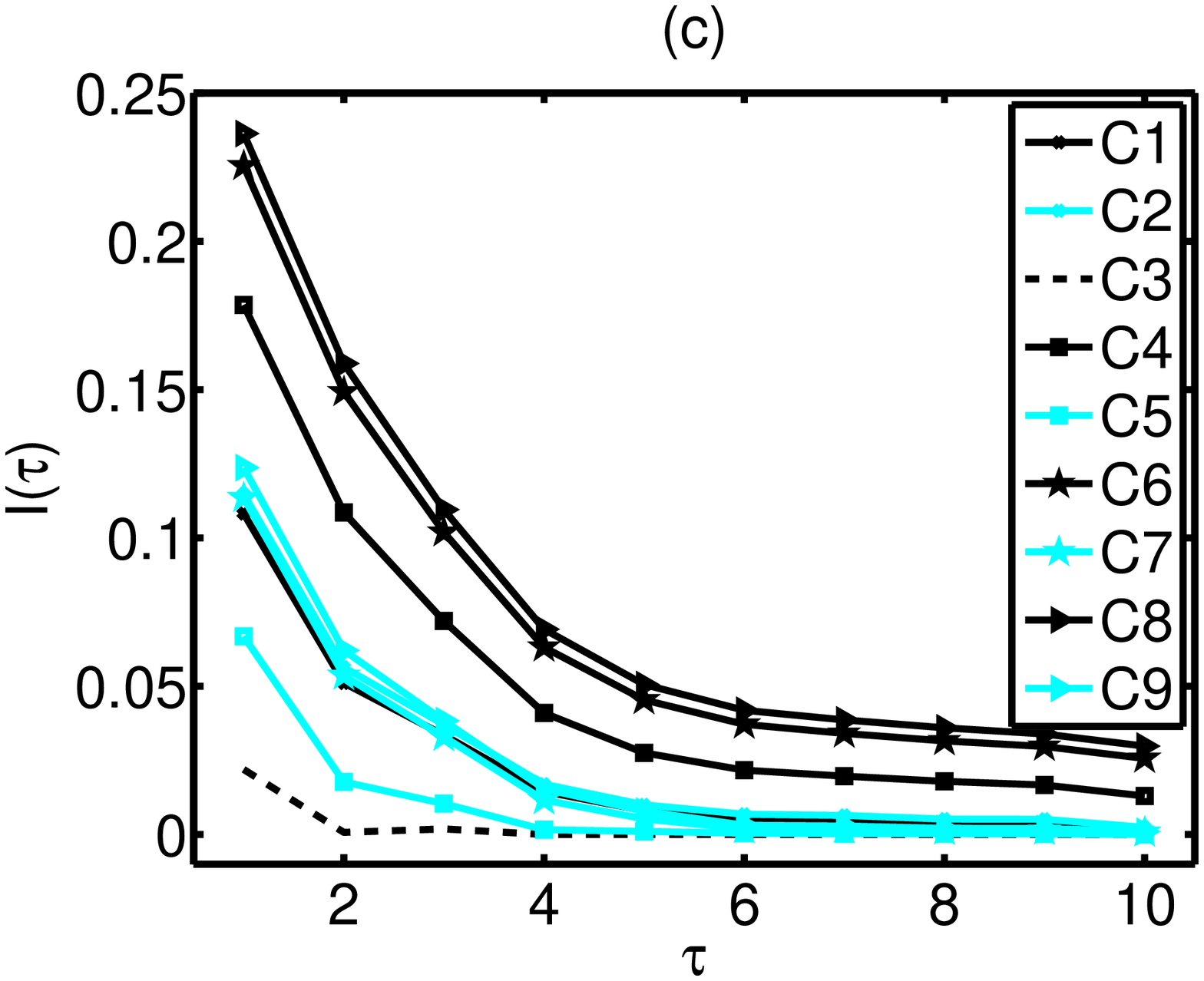}
\includegraphics[height=3.4cm,keepaspectratio]{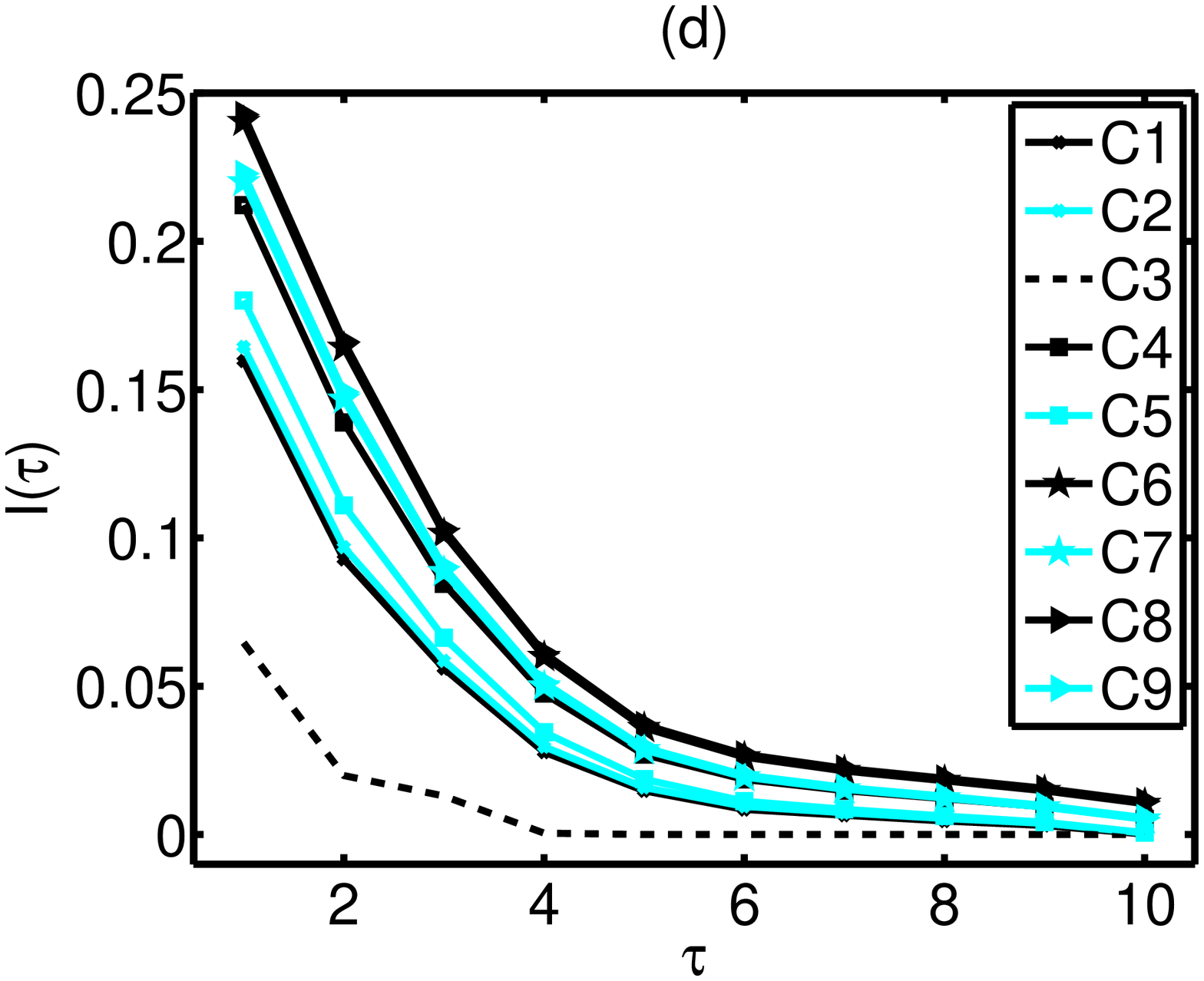}
}}
 \caption{Average of ${\it I}(\tau)$ from the kernel estimator for
lags $\tau =1,\ldots,10$ from $1000$ realizations of the Henon map
from the nine bandwidth selection criteria for noise-free data and
for $n = 512$ in (a) and $n = 8192$ in (b) and for $20\%$ noise
level and $n = 512$ in (c) and $n = 8192$ in (d).}
 \label{fig:Henonkernelcriteria}
\end{figure}

Regarding the 9 criteria selecting $h_1$ (and at cases $h_2$, see
Table~\ref{tab:bandwidth}), the estimated bandwidths vary but
within a small range (for $n=512$ in
Fig.~\ref{fig:Henonkernelcriteria}a they are bounded in [0.1,0.3]
except C3 that always gives larger bandwidths where in this case
is $h_1\simeq0.64$).
%==============================================================================
% Figure9
% =======
% ==============================================================================
All criteria depend on $n$ in a similar way and estimate smaller
bandwidths as $n$ increases giving larger ${\it I}(\tau)$ (see
Fig.~\ref{fig:Henonkernelcriteria}a and b). Thus a kernel estimator
using a specific criterion turns out not to be consistent.

When noise is added to the time series ${\it I}(\tau)$ decreases and
differences with respect to the partitioning parameters are smaller,
as observed in the other estimators. This holds also when a specific
bandwidth selection criterion is used, and in particular the
estimated ${\it I}(\tau)$ is rather stable to the change of $n$ (see
Fig.~\ref{fig:Henonkernelcriteria}c and d).

In the estimation of mutual information with kernels, the range of
bandwidths is usually not searched and a bandwidth is selected
according to a criterion such as the "Gaussian" bandwidth
\cite{Steuer02}. However, our simulations have shown that KE
estimator is strongly dependent on the bandwidth that defines the
partition differently according to the sample size, and these
findings are in agreement with other works \cite{Bonnlander94,
Jones96}.

%========================================================================
\subsection{Evaluation of estimators and their parameters}
%=========================================================================

%In order to compare the estimators, we focused on the optimization
%of the parameters and we were interested in identifying the
%connection between the partition parameter. The usefulness of such
%an investigation is presented by selecting and showing also ${\it
%I}(\tau)$ computed using the standard parameter selection criteria
%from bibliography.
The results on the different estimators have shown a varying
sensitivity of the estimator to its free parameter, where
histogram-based estimators turned out to be the most sensitive.
However, there seems to be a loose correspondence among the
different free parameters $b$, $k$ and $h_1, h_2$ depending also on
the time series length. Thus the differences in the performance of
the estimators can be explained to some degree by the coarseness of
the partition as determined by its free parameter. The choice of $b$
for the histogram-based estimators determines the bin size of the
partition. The analogue of the bin size for the $k$-nearest
neighbors estimator is the size of neighborhoods and for the kernel
estimator is the size of the efficient support for the kernel given
by the bandwidth.

\begin{figure}[h!]
\centerline{\hbox{
\includegraphics[height=3.4cm,keepaspectratio]{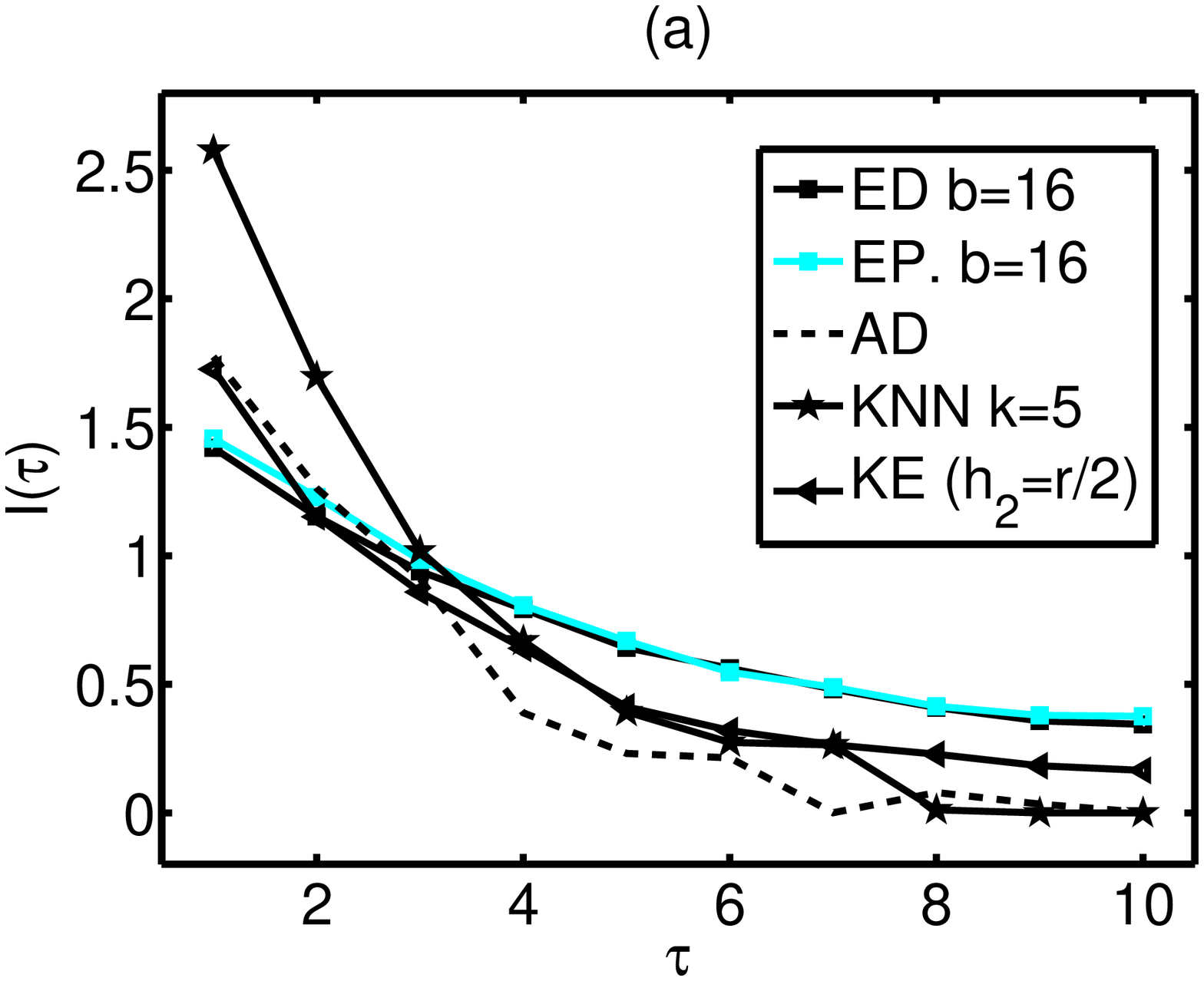}
\includegraphics[height=3.4cm,keepaspectratio]{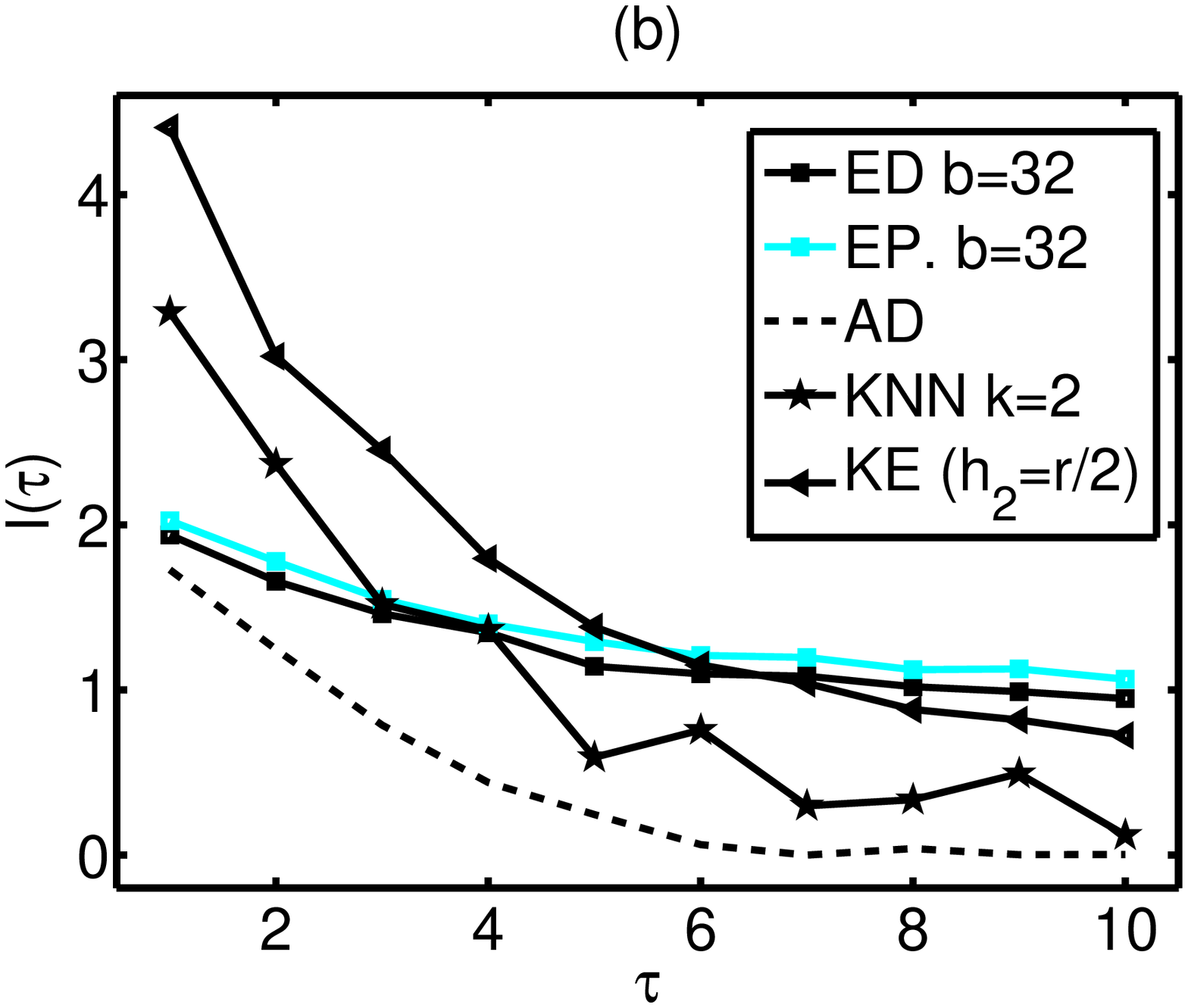}
}}
\centerline{\hbox{
\includegraphics[height=3.4cm,keepaspectratio]{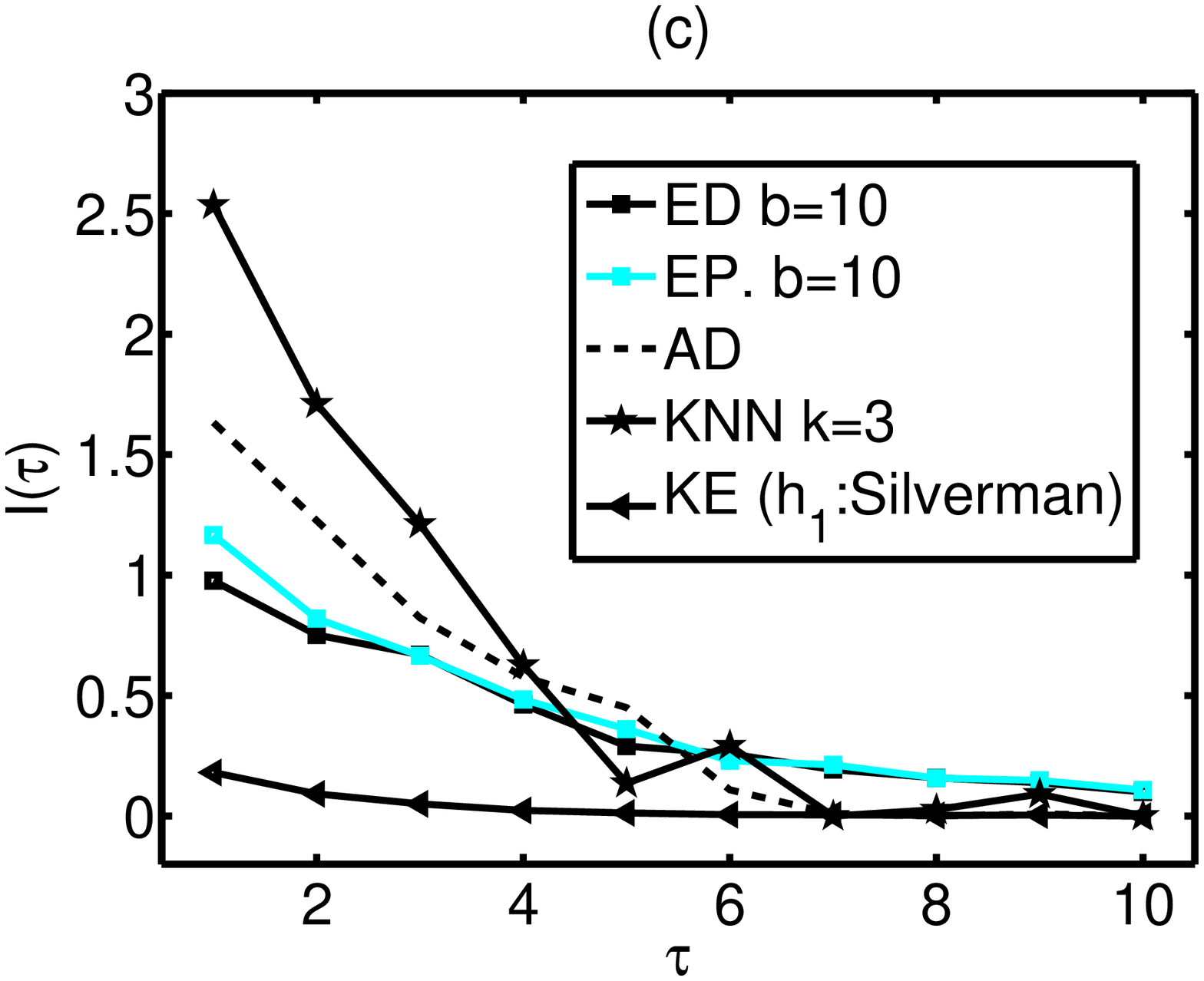}
\includegraphics[height=3.4cm,keepaspectratio]{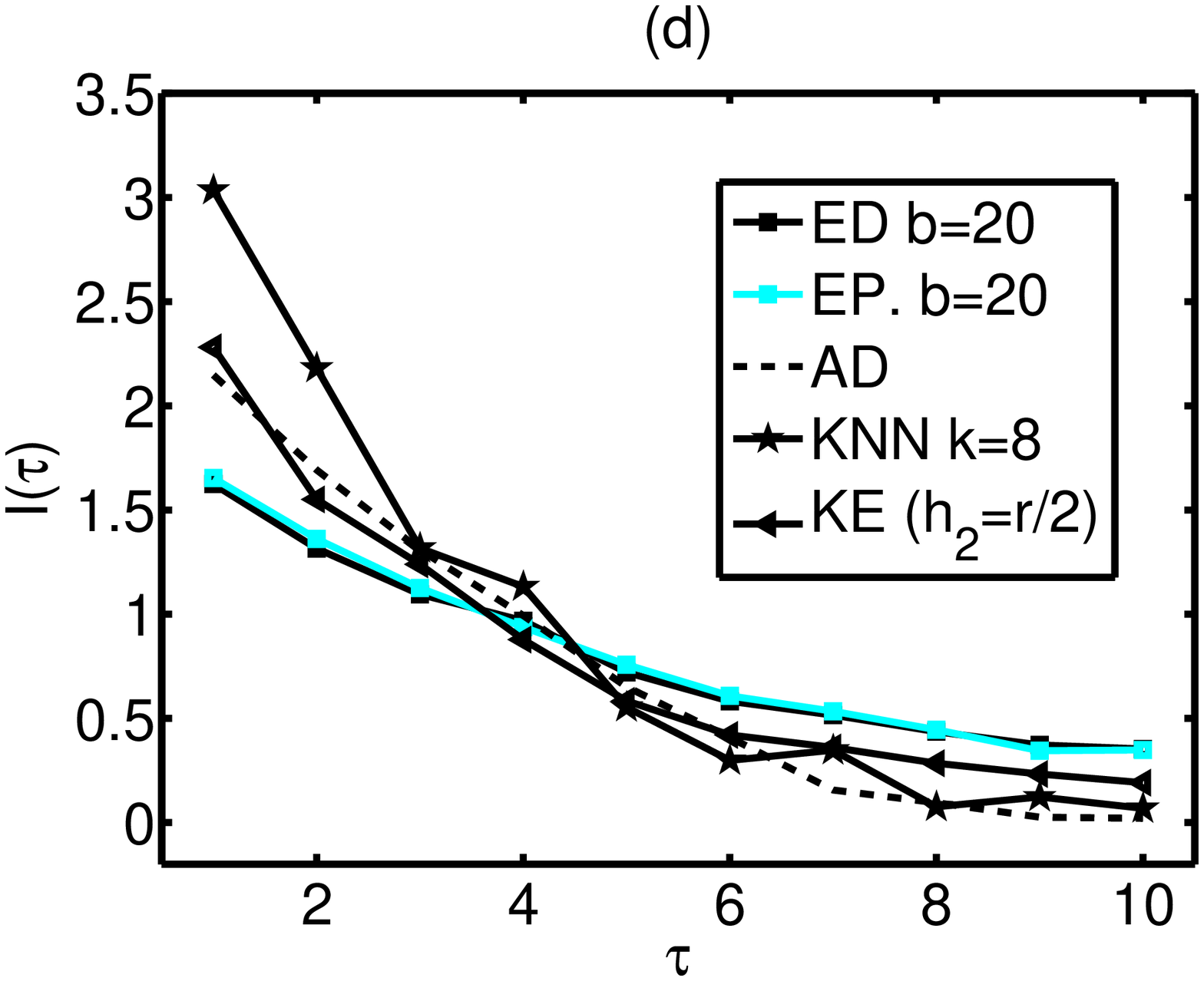}
}}
\caption{${\it I}(\tau)$ from one realization of the Henon map
of from all estimators, for parameters as in the legends for $n =
512$ in (a), (b) and (c) and $n = 1024$ in (d).}
\label{fig:henonallMI}
\end{figure}

In order to check the correspondence of the estimator specific
parameters, first we set a specific bin length $r$ for the
partition given by $b$. For KNN the parameter $k$ is set to the
average number of neighbors within each disc of diameter $r$. For
KE we set $h_2=r/2$. In this way, we attempt to match the
partition for each estimator. However, this selection scheme for
the parameters does not result in similar estimated values of
${\it I}(\tau)$. For example, for a time series of the Henon map
with length $n=512$ when $b=16$ the estimates of ${\it I}(\tau)$
seem to agree at some degree where as when $b=32$ they vary
significantly, as shown in Fig.\ref{fig:henonallMI}.
% ==============================================================================
% Figure10
% =======
%==============================================================================
When the standard criteria for the selection of the parameters are
considered, i.e. $b = \sqrt{n/5}$, $k =3$, and bandwidths
$h_1,h_2$ given by the Silverman's criterion, the estimates of
${\it I}(\tau)$ vary even more (see Fig.\ref{fig:henonallMI}c).
For a bit larger length as $n =1024$ we observed that if we choose
a bit larger value of $b$ as $b = 20$, we get similar estimated
values of ${\it I}(\tau)$(see Fig.\ref{fig:henonallMI}d).

[{\tt to revise this according to the results on differen $n$ and
for the other systems.}] We concluded that the optimization of
parameters is very crucial even more than the choice of the
estimator, as we can see that no estimator exhibits consistency
especially in the case of noise-free data.

It is also important to compare the computational cost of the
estimators. The kernel estimator has the highest computational
cost and the computation time for the histogram-based estimators
is also prohibitive for very large values of $b$. The adaptive
estimator has the advantage of being fast and parameter-free but
tends to give larger ${\it I}(\tau)$ (compared to the other
estimators) for small noise-free time series.

Most of the results of the different estimators are illustrated for
the Henon map in order to facilitate comparisons, but qualitatively
similar results are obtained from the simulations on the Ikeda map
and the Mackey-Glass system.

% ======================
\section{Discussion}
\label{sec:Discussion}
% ======================

Mutual information estimators are not consistent for non-linear
noise-free systems and the choice of parameters is crucial for all
estimators. We cannot find an optimal parameter choice as there is
no consistency. However with addition of noise in the systems, the
choice of the parameters is not that crucial as there is
convergence of the estimated ${\it I}(\tau)$ values and all
estimators seem to be consistent especially for larger time series
lengths. $k$-nearest neighbor estimates of ${\it I}(\tau)$ varies
less with the free parameter ($k$) compared to the other
estimators.

As a general conclusion we can say that all estimators depend on
the parameter choice and therefore is crucial to optimize their
parameter. Also consistency of estimators for linear systems is
not indicative of the estimators behavior for nonlinear systems.
Although consistency of estimators is claimed in many past works
might be due to inclusion of only linear systems in their
evaluation or presence of noise (e.g. real data).

However, this shows that the $k$-nearest neighbor estimator is
computationally more effective when fine partitions are sought,
due to the use of effective data structures in the search for
neighbors.

%==================================================================
\begin{acknowledgments}
This research project is implemented within the framework of the
"Reinforcement Programme of Human Research Manpower" (PENED) and
is co-financed at $90\%$ jointly by E.U.-European Social Fund
($75\%$) and the Greek Ministry of Development-GSRT ($25\%$) and
at $10\%$ by Rikshospitalet, Norway.
\end{acknowledgments}
%================================================================
\newpage

\bibliography{ref}

\end{document}